\documentclass[aps,pre,twocolumn,groupedaddress,amsmath,showpacs,eqsecnum]{revtex4}
   \usepackage{bm}
\usepackage{graphicx,amssymb}
\usepackage{bm}
\usepackage{amsmath}
\newcommand{\beq}{\begin{equation}}
\newcommand{\eeq}{\end{equation}}
\newcommand{\beqa}{\begin{eqnarray}}
\newcommand{\eeqa}{\end{eqnarray}}
\newcommand{\dd}{\text{d}}
\newcommand{\e}{\text{$e$}}
\newcommand{\nn}{\nonumber\\}
\newcommand{\hcs}{\text{hc}}
\newcommand{\tg}{\text{t}}
\newcommand{\Pra}{\text{Pr}}

\begin{document}

\title{System of elastic hard spheres which mimics the transport properties of a granular gas}
\author{Andr\'es Santos}
\email{andres@unex.es}
\homepage{http://www.unex.es/eweb/fisteor/andres/}
\affiliation{Departamento de F\'{\i}sica, Universidad de
Extremadura, E--06071 Badajoz, Spain}
\author{Antonio Astillero}
\email{aavivas@unex.es}
\homepage{http://www.unex.es/eweb/fisteor/antonio/}
\affiliation{Departamento de Inform\'atica, Centro Universitario de
M\'erida, Universidad de Extremadura, E--06800 M\'erida (Badajoz),
Spain}

\date{\today}
\begin{abstract}
The prototype model of a fluidized granular system is a gas of
inelastic hard spheres (IHS) with a constant coefficient of normal
restitution $\alpha$. Using a kinetic theory description we
investigate the two basic ingredients that a model of elastic hard
spheres (EHS) must have in order to mimic the most relevant
transport properties of the underlying IHS gas. First, the EHS gas
is assumed to be subject to the action of an effective drag force
with a friction constant equal to half the cooling rate of the IHS
gas, the latter being evaluated in the local equilibrium
approximation for simplicity. Second, the collision rate of the EHS
gas is reduced by a factor $\frac{1}{2}(1+\alpha)$, relative to that
of the IHS gas. Comparison between the respective Navier--Stokes
transport coefficients shows that the EHS model reproduces almost
perfectly the self-diffusion coefficient and reasonably well the two
transport coefficients defining the heat flux, the shear viscosity
being reproduced within a deviation less than 14\% (for $\alpha\geq
0.5$). Moreover, the EHS model is seen to agree with the fundamental
collision integrals of inelastic mixtures and dense gases. The
approximate equivalence between IHS and EHS is used to propose
kinetic models for inelastic collisions as simple extensions of
known kinetic models for elastic collisions.
\end{abstract}
\pacs{45.70.Mg, 05.20.Dd, 05.60.-k, 51.10.+y }

\maketitle
\section{Introduction\label{sec1}}
As is well known, the prototype model of a granular fluid under
conditions of rapid flow consists of a gas of (smooth)
\textit{inelastic} hard spheres (IHS) characterized by a constant
coefficient of normal restitution $\alpha\leq 1$ \cite{C90,G03}.
When two particles moving with velocities $\mathbf{v}$ and
$\mathbf{v}_1$ collide inelastically, they emerge after collision
with velocities $\mathbf{v}'$ and $\mathbf{v}_1'$, respectively,
given by
\begin{equation}
{\bf v}'={\bf v}-\frac{1+\alpha }{2}({\bf g}\cdot
\widehat{\bm{\sigma}})\widehat{\bm{\sigma}},\quad {\bf v}_{1}'={\bf v}_{1}+\frac{1+\alpha }{2 }({\bf g}\cdot \widehat{\bm{\sigma}
})\widehat{\bm{\sigma}}.  \label{3.3bis}
\end{equation}
Here, $\widehat{\bm{\sigma}}$ is a unit vector directed along the
centers of the two colliding spheres at contact and
$\mathbf{g}\equiv\mathbf{v}-\mathbf{v}_1$ is the pre-collisional
relative velocity. The collision rule (\ref{3.3bis}) conserves
momentum but energy is decreased by a factor proportional to the
degree of inelasticity $1-\alpha^2$, namely
\beq
{v'}^2+{v'_1}^2-v^2-v_1^2=-({\bf g}\cdot
\widehat{\bm{\sigma}})^2\frac{1-\alpha^2}{2}.
\label{n1.10}
\eeq
Stated differently, while the component of the relative velocity
orthogonal to $\widehat{\bm{\sigma}}$ does not change upon
collision, the magnitude of the parallel component decreases by a
factor $\alpha$: ${\bf g}'\cdot \widehat{\bm{\sigma}}=-\alpha {\bf
g}\cdot \widehat{\bm{\sigma}}$, where
$\mathbf{g}'\equiv\mathbf{v}'-\mathbf{v}_1'$ is the post-collisional
relative velocity. {}From Eq.\ (\ref{3.3bis}) it is straightforward
to get the pre-collisional or restituting velocities ${\bf v}''$ and
${\bf v}_1''$ giving rise to post-collisional velocities ${\bf v}$
and ${\bf v}_1$:
\begin{equation}
{\bf v}''={\bf v}-\frac{1+\alpha }{2\alpha}({\bf g}\cdot
\widehat{\bm{\sigma}})\widehat{\bm{\sigma}},\quad {\bf v}_{1}''={\bf
v}_{1}+\frac{1+\alpha }{2\alpha}({\bf g}\cdot \widehat{\bm{\sigma}
})\widehat{\bm{\sigma}}, \label{3.3}
\end{equation}
where $\mathbf{g}=\mathbf{v}-\mathbf{v}_1$ is now the
post-collisional relative velocity.

The inelasticity of collisions contributes to a decrease of the granular temperature $T(t)$
(proportional to the mean kinetic energy per particle in the Lagrangian frame), i.e.,
\begin{equation}
\left.\frac{\partial T}{\partial t}\right|_{\text{coll}}=-\zeta T,
\label{n1.1}
\end{equation}
where $\zeta\sim \nu( 1-\alpha^2)$ is the \textit{cooling rate},
$\nu$ being an effective collision frequency. Equation (\ref{n1.1})
implies that in order to reach a steady state an external energy
input is needed to compensate for the collisional cooling.

In the case of a gas of \textit{elastic} hard spheres (EHS), energy
is conserved by collisions. However,  a cooling effect can be
generated by the application of a drag force proportional to the
particle velocity, i.e.,
\begin{equation}
\left.\frac{\partial T}{\partial t}\right|_{\text{drag}}=-2\gamma T,
\label{n1.2}
\end{equation}
where $\gamma$ is the friction coefficient \cite{note0}. Obviously,
the choice $\gamma=\frac{1}{2}\zeta$ makes the drag force produce
the same cooling effect on the EHS system as inelasticity does on
the IHS one. As a consequence, at a macroscopic level of
description, the hydrodynamic balance equations of mass, momentum,
and energy for the IHS gas are (formally) identical to those for the
frictional EHS gas:
\beq
D_t n+n\nabla\cdot \mathbf{u}=0,
\label{n1.3}
\eeq
\beq
D_t\mathbf{u}+\frac{1}{mn}\nabla\cdot\mathsf{P}=\mathbf{0},
\label{n1.4}
\eeq
\beq
D_tT+\frac{2}{dn}\left(\nabla\cdot\mathbf{q}+\mathsf{P}:\nabla
\mathbf{u}\right)=-\zeta T.
\label{n1.5}
\eeq
Here, $D_t\equiv\partial_t+\mathbf{u}\cdot\nabla$ is the material
time derivative, $d$ is the dimensionality of the system, $m$ is the
mass of a sphere, $n$ is the number density, $\mathbf{u}$ is the
flow velocity, $T$ is the granular temperature, $\mathsf{P}$ is the
pressure tensor, and $\mathbf{q}$ is the heat flux. The right-hand
side term in Eq.\ (\ref{n1.5}) comes from Eq.\ (\ref{n1.1}) in the
case of IHS, whereas it comes from Eq.\ (\ref{n1.2}) (with
$\gamma=\frac{1}{2}\zeta$) in the case of EHS.

\begin{figure}[h]
 \includegraphics[width=.90 \columnwidth]{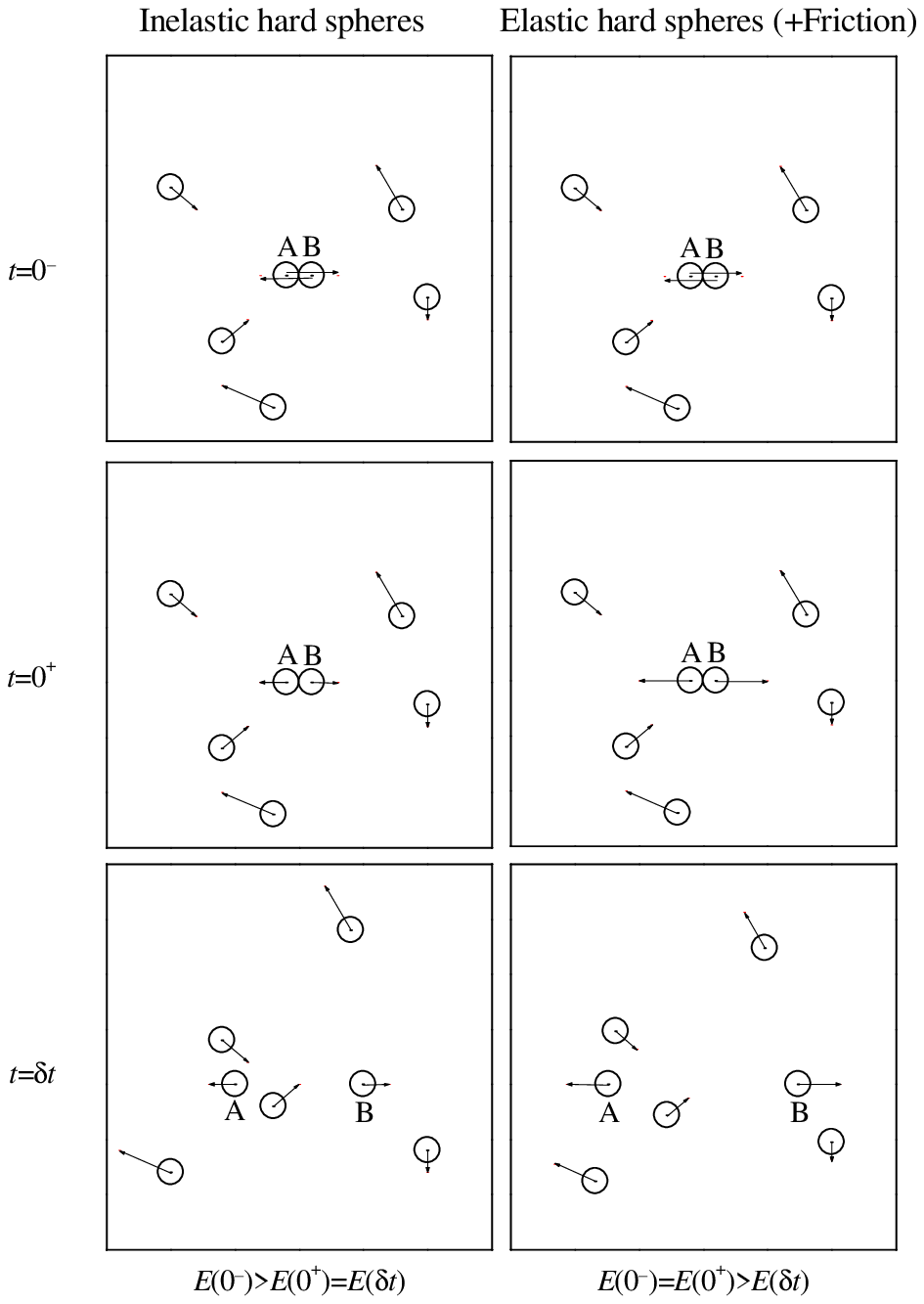}
 \caption{Sketch of the microscopic dynamics of inelastic hard spheres (IHS, left panel)
 and elastic hard spheres  under a friction force (EHS, right panel). At $t=0^-$ both systems
 are prepared in the same microstate and particles A and B are about to collide.
 In the IHS system, particles A and B recede immediately after collision (at $t=0^+$)
 with a  relative velocity smaller than the pre-collision one, so the mean kinetic
 energy decreases, $E(0^+)<E(0^-)$. During the time step $\delta t$ no collision
 takes place and the particles move ballistically, so $E(\delta t)=E(0^+)$.
 In the EHS system, the collision between particles A and B is elastic,
 so $E(0^+)=E(0^-)$. However, during the time step $\delta t$ all the
 particles feel the drag force and, consequently, $E(\delta t)<E(0^+)$.
 Note that at $t=\delta t$ the respective microstates in the IHS and EHS
 systems are different, even if the energy loss $E(0^-)-E(\delta t)$
 is the same.\label{sketch}}
 \end{figure}
Despite the trivial common structure of the macroscopic balance
equations (\ref{n1.3})--(\ref{n1.5}) for the free IHS and the driven
EHS gases, the underlying microscopic dynamics is physically quite
different in both systems: in the IHS gas (i) each colliding pair
loses energy upon collision but (ii) all the particles move freely
between two successive collisions; in the EHS case, however, (i)
energy is conserved by collisions but (ii) all the particles lose
energy between collisions due to the action of the drag force.
Therefore, during a certain small time step, only a small fraction
of particles is responsible for the cooling of the system in the IHS
case, whereas all the particles contribute to the cooling in the EHS
case. These differences are sketched in Fig.\ \ref{sketch}. In
principle, there is no reason  to expect that the relevant
nonequilibrium physical properties (e.g., the one-particle velocity
distribution function) are similar for IHS and frictional EHS under
the same conditions.

Let us consider, for instance, the homogeneous cooling state
\cite{vNE98}. In that case, the energy balance equation (\ref{n1.5})
becomes
\beq
\frac{\dd T_\hcs(t)}{\dd t}=-\zeta_\hcs (t) T_\hcs(t).
\label{n1.6}
\eeq
The solution to Eq.\ (\ref{n1.6}) is Haff's law:
\beq
T_\hcs(t)=\frac{T_\hcs(0)}{\left[1+\frac{1}{2}\zeta_\hcs(0)t\right]^2},
\label{n1.7}
\eeq
where we have taken into account that $\zeta_\hcs(t)\propto
T_\hcs^{1/2}(t)$. The above cooling law is valid for the homogeneous
cooling state of both IHS and frictional EHS, provided that in the
latter one considers a time-dependent friction coefficient
$\gamma\propto T^{1/2}$ and $\gamma(0)=\zeta_\hcs(0)$. The
homogeneous Boltzmann equation admits in both cases a scaling
solution of the form
\beq
\begin{array}{c}
f_\hcs(\mathbf{v},t)=n\left[{m}/{2T_\hcs(t)}\right]^{d/2}
f_\hcs^*\left(\mathbf{c}(t)\right),\\
\mathbf{c}(t)=\mathbf{v}/\sqrt{2T_\hcs(t)/m}.
\end{array}
\label{n1.8}
\eeq
On the other hand,  while the distribution function is a Gaussian
$f_\hcs^*(\mathbf{c})=\pi^{-d/2}\e^{-c^2}$ for EHS \cite{GSB90},
deviations from a Gaussian are present in the case of IHS
\cite{vNE98,BRC96,MS00}. These deviations are measured, for
instance, by a nonzero fourth cumulant (or kurtosis),
\beq
a_2\equiv\frac{4}{d(d+2)}\langle c^4\rangle -1,
\label{n1.9}
\eeq
and by an overpopulated high-energy tail $f_\hcs^*(\mathbf{c})\sim
\e^{-A c}$. In the case of a gas heated by a white-noise forcing,
the steady-state distribution function is again a Gaussian for
frictional EHS, while $a_2\neq 0$ and $f^*(\mathbf{c})\sim \e^{-A
c^{3/2}}$ for IHS \cite{vNE98,MS00}.

The above two examples are sufficient to illustrate that, obviously,
the IHS and driven EHS systems are not strictly equivalent. On the
other hand, the differences between the homogeneous solutions for
IHS and EHS are not quantitatively important in the domain of
\textit{thermal} speeds (for instance, $|a_2|\lesssim 0.02$ for IHS
with $\alpha\gtrsim 0.7$). Therefore, it is still possible that both
systems exhibit comparable departures from equilibrium in
\textit{inhomogeneous} states, in which case transport of momentum
and/or energy are the most relevant phenomena. As a matter of fact,
one of the most distinctive features of granular gases, namely the
clustering instability, has a hydrodynamic origin \cite{G03,BRC96},
and so it must also appear in a gas of EHS under the action of a
drag force with a state-dependent friction coefficient $\gamma$
proportional to a characteristic local collision frequency.

The aim of this paper is to investigate to what extent a frictional
EHS gas can ``disguise'' as an IHS gas in what concerns the
transport properties of mass, momentum, and/or energy, which are
dominated by the thermal domain of the velocity distribution
function. If that were the case, the significant body of work
already available for the kinetic theory of normal gases could be
exploited to provide a practical tool for granular gases. A
preliminary report of this work has been given in Ref.\ \cite{AS04}.
In Sec.\ \ref{sec2} we construct a minimal model of EHS in the
framework of the Boltzmann equation that intends to capture the
basic properties of the Boltzmann equation for IHS with a given
coefficient of restitution $\alpha$. First, as discussed above, the
EHS system is assumed to be under the influence of a drag force. By
equating the right-hand sides of Eqs.\ (\ref{n1.1}) and
(\ref{n1.2}), the friction coefficient $\gamma$ of the EHS gas
should be chosen as half  the cooling rate  $\zeta(\alpha)$ of the
true IHS gas. However, this is not practical since $\zeta$ is a
functional of the nonequilibrium velocity distribution of IHS and we
want an \textit{autonomous} Boltzmann equation for EHS, i.e., an
equation that does not require the previous knowledge of the
solution of the Boltzmann equation for IHS. One could autonomously
define $\gamma$ as the same functional of the EHS distribution as
$\zeta$ is of the IHS distribution, but this would also result in a
too complicated model. For these reasons, we choose
$\gamma=\frac{1}{2}\zeta_0(\alpha)$, where $\zeta_0$ is the cooling
rate in the local equilibrium approximation. This is consistent with
the fact that the velocity distribution function of EHS is a
Gaussian in the homogeneous cooling state, as well as in the steady
state driven by a white-noise thermostat. The price to be paid by
the simple choice $\gamma=\frac{1}{2}\zeta_0$ is that the
equivalence between the respective sources of cooling, Eqs.\
(\ref{n1.1}) and (\ref{n1.2}), is only approximate. As a second
ingredient of the model, the collision rate of the EHS gas, relative
to that of the IHS gas, defines a dimensionless parameter $\beta$
that can be freely chosen to optimize the agreement with the IHS
properties. In order to have a clue on an optimal choice of $\beta$,
the Navier--Stokes transport coefficients of both systems are
compared in Sec.\ \ref{sec3}. As a compromise between simplicity and
accuracy, we simply choose $\beta(\alpha)=\frac{1}{2}(1+\alpha)$.
The (approximate) mapping EHS$\rightarrow$IHS allows one to extend
directly to granular gases those kinetic models originally proposed
for conventional gases \cite{C88,GS03}. This is illustrated in Sec.\
\ref{sec7} for the Bhatnagar--Gross--Krook \cite{BGK54} and the
ellipsoidal statistical \cite{H66} kinetic models. The extension of
the approximate equivalence between inelastic and (frictional)
elastic particles to mixtures, dense gases, and Maxwell models is
discussed in Appendices \ref{sec5.1}--\ref{sec5.3}, respectively. In
particular, the applications to mixtures and to dense gases
reinforce the choice $\beta(\alpha)=\frac{1}{2}(1+\alpha)$. The
paper ends with some concluding remarks in Sec.\ \ref{sec8}.

\section{Model of frictional elastic hard spheres\label{sec2}}
\subsection{Basic properties of the Boltzmann equation for inelastic hard spheres}
The Boltzmann equation for a gas of inelastic hard spheres (IHS) is
\cite{GS95,BDS97,vNE01}
\begin{equation}
\left( \partial _{t}+{\bf v\cdot \nabla }\right) f=J^{(\alpha)}[f,f],
\label{3.1}
\end{equation}
where  $f({\bf r},{\bf v};t)$ is the one-particle velocity distribution function and $J^{(\alpha)}[f,f]$ is the Boltzmann collision operator
\begin{eqnarray}
J^{(\alpha)}[f,f]=\sigma ^{d-1}\int \dd{\bf v}_{1}\int \dd\widehat{\bm{\sigma}}\,\Theta (
{\bf g}\cdot \widehat{\bm{\sigma}})({\bf g}\cdot \widehat{\bm{\sigma}})\nonumber\\
\times\left[ \alpha ^{-2}f({\bf v}'')f({\bf v}
_{1}'')-f({\bf v})f({\bf v}_{1})\right] ,
\label{3.2}
\end{eqnarray}
where the explicit dependence of $f$ on ${\bf r}$ and $t$ has been
omitted. In Eq.\ (\ref{3.2}), $\sigma$ is the diameter of a sphere
and $\Theta $ is the Heaviside step function.
 The pre-collisional or
restituting velocities ${\bf v}''$ and ${\bf v}_1''$ are
given by Eq.\ (\ref{3.3}).
Of course, the collision operator for elastic hard spheres (EHS), $J^{(1)}[f,f]$,
is obtained from Eqs.\ (\ref{3.2}) and (\ref{3.3}) by simply setting $\alpha=1$.

The first $d+2$ moments of the velocity distribution function define the number density
\begin{equation}
n(\mathbf{r},t)=\int \dd\mathbf{v}\, f({\bf r},{\bf v};t),
\label{b1}
\end{equation}
the nonequilibrium flow velocity
\begin{equation}
\mathbf{u}(\mathbf{r},t)=\frac{1}{n(\mathbf{r},t)}\int \dd\mathbf{v}\, \mathbf{v} f({\bf r},{\bf v};t)=\langle \mathbf{v}\rangle,
\label{b2}
\end{equation}
and the \textit{granular} temperature
\begin{equation}
T(\mathbf{r},t)=\frac{m}{n(\mathbf{r},t)d}\int \dd\mathbf{v}\, V^2(\mathbf{r},t)f({\bf r},{\bf v};t)=\frac{m\langle V^2\rangle}{d} ,
\label{b3}
\end{equation}
where ${\bf V}({\bf r},t)\equiv{\bf v}-{\bf u}({\bf r},t)$ is the
peculiar velocity. The basic properties of $J^{(\alpha)}[f,f]$  are
those that determine the form of the macroscopic balance equations
for mass, momentum, and energy,
\begin{equation}
\int \dd{\bf v}\left(
\begin{array}{c}
1 \\
m{\bf v} \\
\frac{1}{2}mV^{2}
\end{array}
\right) J^{(\alpha)}[f,f]=\left(
\begin{array}{c}
0 \\
{\bf 0} \\
-\frac{d}{2}nT\zeta
\end{array}
\right) .  \label{3.4}
\end{equation}
By standard manipulations of the collision operator, the cooling
rate can be written as \cite{BDS97,BDS99}
\beq
\zeta(\mathbf{r},t) =(1-\alpha ^{2})\frac{m\pi
^{\frac{d-1}{2}}\sigma ^{d-1}}{4d\Gamma \left(
\frac{d+3}{2}\right)}\frac{n(\mathbf{r},t)}{T(\mathbf{r},t)}\langle
V_{12}^3\rangle,
\label{3.5}
\eeq
where
\beq
\langle V_{12}^3\rangle =\frac{1}{n^2(\mathbf{r},t)} \int \dd{\bf
v}_1\int \dd{\bf v}_{2}\,|{\bf v }_1-{\bf v}_{2}|^{3}f({\bf r},{\bf
v}_1;t)f({\bf r},{\bf v}_{2};t)
\label{3.5bis}
\eeq
is the average value of the cube of the relative speed. The
properties (\ref{3.4}) lead to the balance equations
(\ref{n1.3})--(\ref{n1.5}) with the following kinetic expressions
for the pressure tensor and the heat flux,
\begin{equation}
P_{ij}=m\int \dd\mathbf{v}\,  V_i V_j f(\mathbf{v}),
\label{c1}
\end{equation}
\begin{equation}
\mathbf{q}=\frac{m}{2}\int \dd\mathbf{v}\, V^2 \mathbf{V} f(\mathbf{v}).
\label{c2}
\end{equation}

The cooling rate $\zeta$ is a nonlinear functional of the
distribution function $f$ through the average $\langle
V_{12}^3\rangle$. As a consequence, $\zeta$ cannot be explicitly
evaluated unless the Boltzmann equation is solved. Nevertheless, a
simple \textit{estimate}  is obtained from Eq.\ (\ref{3.5bis}) by
replacing the actual distribution function $f$ by the \textit{local
equilibrium} distribution
\begin{equation}
f_0(\mathbf{v})=n(m/2\pi T)^{d/2}\exp(-mV^2/2T).
\label{3.5.1}
\end{equation}
In that case,
\beqa
\langle V_{12}^3\rangle&\to& \langle V_{12}^3\rangle_0=2^{3/2}
\langle V^3\rangle_0\nn
&=&4\pi^{-d/2}\Omega_d\Gamma\left(\frac{d+3}{2}\right)\left(\frac{T}{m}\right)^{3/2},
\label{n2.1}
\eeqa
where $\Omega_d\equiv 2\pi^{d/2}/\Gamma(d/2)$ is the total solid
angle. When the approximation (\ref{n2.1}) is inserted into Eq.\
(\ref{3.5}), one gets the local equilibrium cooling rate
 \cite{BDS99,BDKS98}
\begin{equation}
\zeta_0(\mathbf{r},t)=\zeta_0^*\nu_0(\mathbf{r},t), \quad \zeta_0^*\equiv \frac{d+2}{4d}
(1-\alpha ^{2}), \label{3.14}
\end{equation}
where
\begin{equation}
\nu_0=
\frac{4\Omega_d}{\sqrt{\pi}(d+2)}n\sigma ^{d-1}\left( \frac{T}{m}
\right) ^{1/2} \label{3.14b}
\end{equation}
is an effective collision frequency. The local equilibrium estimate
(\ref{3.14}) expresses the cooling rate as a functional of $f$
through the local density and temperature only. In addition, its
dependence on inelasticity is simply $\zeta_0\propto 1-\alpha^2$.

It is now convenient to introduce the \textit{modified} collision operator \cite{BDS99}
\begin{equation}
\bar{J}^{(\alpha)}[f,f]\equiv J^{(\alpha)}[f,f]-\frac{\zeta }{2}\frac{\partial }{\partial
{\bf v}}\cdot \left( {\bf V}f\right) .  \label{3.9}
\end{equation}
By construction, the operator $\bar{J}^{(\alpha)}[f,f]$ has the properties
\begin{equation}
 \int \dd{\bf v}\left(
\begin{array}{c}
1 \\
m{\bf v} \\
\frac{1}{2}mV^{2}
\end{array}
\right) \bar{J}^{(\alpha)}[f,f]=\left(
\begin{array}{c}
0 \\
{\bf 0} \\
0
\end{array}
\right) ,  \label{3.10}
\end{equation}
as follows from direct evaluation using Eq.\ (\ref{3.4}).

\subsection{The model}
The modified inelastic collision operator $\bar{J}^{(\alpha)}[f,f]$
shares with the elastic collision operator $J^{(1)}[f,f]$  the
property of having $d+2$ vanishing low velocity moments. Of course,
$\bar{J}^{(\alpha)}[f,f]$ and $J^{(1)}[f,f]$ differ in many other
aspects. For instance, in the homogeneous cooling state the velocity
distribution function is the solution to
$\bar{J}^{(\alpha)}[f_\hcs,f_\hcs]=0$ \cite{BDS99}, which differs
from a Gaussian, the latter being the solution to
${J}^{(1)}[f,f]=0$. Moreover, the elastic collision operator
satisfies the H-theorem, namely
\beq
\int\dd \mathbf{v}(\ln f) {J}^{(1)}[f,f]\leq 0,
\label{n2.2}
\eeq
while an H-theorem has not been proven for
$\bar{J}^{(\alpha)}[f,f]$. Despite these differences, the common
properties (\ref{3.10}) suggest the possibility that the operators
$\bar{J}^{(\alpha)}[f,f]$ and ${J}^{(1)}[f,f]$ have a similar
behavior in the domain of thermal speeds (i.e., for $V\lesssim
2\sqrt{2T/m}$). This expectation  can be exploited to propose the
approximation
\begin{equation}
\bar{J}^{(\alpha)}[f,f]\to \beta J^{(1)}[f,f],
\label{b3bis}
\end{equation}
where $\beta(\alpha)$ is a positive constant to be determined later
on. Its introduction does not invalidate Eq.\ (\ref{3.10}) but
allows us to fine-tune the approximate equivalence between both
operators. In agreement with the spirit of the above discussion we
further approximate the true cooling rate given by Eq.\ (\ref{3.5})
by the local equilibrium estimate (\ref{3.14}). In summary, our
model consists of the replacement
\begin{equation}
J^{(\alpha)}[f,f]\to \beta J^{(1)}[f,f]+\frac{\zeta_0 }{2}\frac{\partial }{\partial
{\bf v}}\cdot \left( {\bf V}f\right),
\label{b4}
\end{equation}
so that the Boltzmann equation (\ref{3.1}) becomes
\begin{equation}
\left( \partial _{t}+{\bf v\cdot \nabla }-\frac{\zeta_0 }{2}\frac{\partial }{\partial
{\bf v}}\cdot {\bf V}\right) f=\beta J^{(1)}[f,f].
\label{n2.3}
\end{equation}

In this model, the gas of \textit{inelastic} hard spheres with a
given coefficient of restitution $\alpha$ is replaced by an
``equivalent'' gas of \textit{elastic} hard spheres subject to the
action of a drag force $\mathbf{F}_{\text{drag}}=-m\gamma\mathbf{V}$
with $\gamma=\frac{1}{2}\zeta_0$.   While this drag force does not
affect the conservation of momentum \textit{on average}, this is not
 so at a \textit{microscopic} level. To clarify this
point, let us consider  the particles inside a small box
$\delta\mathbf{r}$ centered about the point $\mathbf{r}$ at time $t$
in a given microstate. Because of the action of the drag force, the
velocities of those particles change during a short time interval
$\delta t$ as
\beq
\mathbf{v}_i(t)\to \mathbf{v}_i(t+\delta
t)=\mathbf{v}_i(t)-\gamma(\mathbf{r},t)\left[\mathbf{v}_i(t)-\mathbf{u}(\mathbf{r},t)\right]\delta
t.
\eeq
Summing over all the $\delta N$ particles inside the box, we find
that, in general, $\sum_{i\in\delta\mathbf{r}}\mathbf{v}_i(t+\delta
t)\neq \sum_{i\in\delta\mathbf{r}}\mathbf{v}_i(t)$ since
$\sum_{i\in\delta\mathbf{r}}\mathbf{v}_i(t)\neq \delta N
\mathbf{u}(\mathbf{r},t)$. On the other hand, the momentum is
conserved when averaging over all the microstates, i.e.,
$\langle\sum_{i\in\delta\mathbf{r}}\mathbf{v}_i(t+\delta
t)\rangle=\langle
\sum_{i\in\delta\mathbf{r}}\mathbf{v}_i(t)\rangle$, where we have
taken into acoount that, by definition,
$\mathbf{u}(\mathbf{r},t)=\langle
\sum_{i\in\delta\mathbf{r}}\mathbf{v}_i(t)\rangle/\langle\delta
N\rangle$. This averaging process is already built in the Boltzmann
equation (\ref{n2.3}).

The fact that in the friction constant the actual cooling rate
$\zeta$ of the granular gas is approximated by $\zeta_0$ (or,
equivalently, $\langle V_{12}^3\rangle\to\langle V_{12}^3\rangle_0$)
is dictated by simplicity  since it does not seem necessary to
retain the detailed functional dependence of Eq.\ (\ref{3.5bis})
when on the other hand the coarse-grained approximation
(\ref{b3bis}) is being used. In other words, the discrepancies due
to $\zeta\to \zeta_0$  may be expected to be less important than
those associated with the approximation (\ref{b3bis}) itself. In any
case, if one wishes to keep the true cooling rate $\zeta$ in the
model (\ref{b4}), $\zeta$ must be interpreted as a functional of the
solution of  Eq.\ (\ref{n2.3}) itself, not as a functional of the
solution  of Eq.\ (\ref{3.1}). Otherwise, Eq.\ (\ref{n2.3}) would
not be an autonomous equation and it would be necessary to know the
solution of the IHS Boltzmann equation (\ref{3.1})  before dealing
with the EHS Boltzmann equation (\ref{n2.3}), what is not only
impractical but artificially complicated as well.

In Eq.\ (\ref{3.2}) the coefficient of restitution $\alpha$ appears
both explicitly (by the factor $\alpha^{-2}$ inside the collision
integral) and implicitly [through the collision rule (\ref{3.3})].
In contrast, in the model (\ref{b4}) $\alpha$ appears only
explicitly, as well as outside the collision integral, through the
approximate cooling rate $\zeta_0\propto 1-\alpha^2$ and the
parameter $\beta(\alpha)$ still to be determined. This
simplification can be justified as long as one is mainly interested
in the \textit{gross} effects of inelasticity on the nonequilibrium
velocity distribution function, while the \textit{fine} details
might not be captured by the model (\ref{b4}). However, as seen in
the companion paper \cite{AS05}, the reliability of the EHS model
turns out to be higher than the expected one.

In principle, there is no reason to expect that the gas of EHS which
most efficiently mimics the relevant transport properties of the
granular gas is made of particles with a diameter $\sigma'$ equal to
the diameter $\sigma$ of the inelastic spheres. If $\sigma'$ and
$\sigma$ were equal, then both systems would have the same mean free
time but not necessarily the same rate of momentum and energy
transfer upon collisions. The  effect associated with
$\sigma'\neq\sigma$ is accounted for by the parameter $\beta$. Since
$J^{(1)}[f,f]$ is defined by setting $\alpha=1$ in Eq.\ (\ref{3.2}),
it is proportional to $\sigma^{d-1}$, not to ${\sigma'}^{d-1}$.
Therefore, $\beta J^{(1)}[f,f]$ is the collision operator of EHS of
diameter $\sigma'=\beta^{1/(d-1)}\sigma$. Alternatively, Eq.\
(\ref{n2.3}) can be rewritten as
\begin{equation}
\left( \partial _{t'}+{\bf v\cdot \nabla' }-\frac{\zeta_0 }{2\beta}\frac{\partial }{\partial
{\bf v}}\cdot {\bf V}\right) f=J^{(1)}[f,f],
\label{n2.3bis}
\end{equation}
where $t'\equiv \beta t$ and
$\nabla'=\partial/\partial{\mathbf{r}'}$ with
$\mathbf{r}'\equiv\beta\mathbf{r}$. According to Eq.\
(\ref{n2.3bis}), the original IHS system is replaced by an EHS
system with the same diameter $\sigma'=\sigma$, but with a friction
constant $\gamma'=\zeta_0/2\beta$ and spatial and temporal variables
scaled by a factor $\beta$ with respect to those of the IHS system.
In what follows, we will use for the model the form (\ref{n2.3})
rather than the form (\ref{n2.3bis}) and will view $\beta$ as a
correction factor to modify the collision rate of the equivalent
system of EHS, relative to the collision rate of the IHS system.
This implies that after a certain common time interval $\Delta t$
the number of collisions experienced by the EHS is in general
different from that of the IHS.   In the next Section we make a
definite proposal for $\beta(\alpha)$ based on the comparison
between the EHS and IHS Navier--Stokes transport coefficients.

\section{Navier--Stokes transport coefficients\label{sec3}}
\subsection{Stress tensor and heat flux}
The irreversible momentum and energy transport are measured by the
stress tensor $\Pi_{ij}=P_{ij}-p\delta_{ij}$ (where
$p=nT={d}^{-1}\text{Tr}\,\mathsf{P}$ is the hydrostatic pressure)
and the heat flux $\mathbf{q}$. By an extension of the
Chapman--Enskog method \cite{CC70} to the case of inelastic
collisions \cite{BDKS98,GD99,BC01,GM02,S03}, one gets the
Navier--Stokes constitutive equations
\begin{equation}
\Pi_{ij}=-\eta \left(\nabla_i u_j+\nabla_j u_i-\frac{2}{d}\nabla\cdot\mathbf{u}\,\delta_{ij}\right),
\label{c3}
\end{equation}
\begin{equation}
\mathbf{q}=-\lambda \nabla T-\mu \nabla n,
\label{c4}
\end{equation}
where $\eta$ is the shear viscosity, $\lambda$ is the thermal
conductivity,  and $\mu$ is a transport coefficient with no
counterpart in the elastic case. For IHS, the explicit expressions
for the transport coefficients in the \textit{first Sonine
approximation} are given by \cite{BDKS98,BC01}
\begin{equation}
\eta=\frac{nT}{\nu_0}\frac{1}{\nu_\eta^*-\frac{1}{2}\zeta^*},
\label{c5}
\end{equation}
\begin{equation}
\lambda=\frac{nT}{m\nu_0}\frac{d+2}{2}\frac{1+2a_2^\hcs}{\nu_\lambda^*-2\zeta^*},
\label{c6}
\end{equation}
\begin{equation}
\mu=\frac{T^2}{m\nu_0}\frac{d+2}{2}\frac{\zeta^*+a_2^\hcs\nu_\lambda^*}{(\nu_\lambda^*-\frac{3}{2}\zeta^*)(\nu_\lambda^*-2\zeta^*)}.
\label{c7}
\end{equation}
In these equations, the effective collision frequency $\nu_0$ is defined by Eq.\ (\ref{3.14b}),
\begin{equation}
a_2^\hcs=\frac{16(1-\alpha)(1-2\alpha^2)}{9+24d-\alpha(41-8d)
+30\alpha^2(1-\alpha)}
\label{c8}
\end{equation}
is an estimate of the kurtosis of the velocity distribution function
[cf.\ Eq.\ (\ref{n1.9})] in the homogeneous cooling state
\cite{vNE98,note}, and
\begin{equation}
\zeta^*=\zeta_0^*\left(1+\frac{3}{16}a_2^\hcs\right)
\label{c9}
\end{equation}
is the (reduced) cooling rate in the same state. Moreover,
\begin{eqnarray}
\nu_\eta^*&=&\frac{\int \dd\mathbf{v}\,\mathsf{D}:{\cal L}^{(\alpha)}f_0\mathsf{D}}{\nu_0\int \dd\mathbf{v}\,f_0\mathsf{D}:\mathsf{D}}\nonumber\\
&=&\frac{3}{4d}\left(1-\alpha+\frac{2}{3}d\right)(1+\alpha)\left(1
-\frac{1}{32}a_2^\hcs\right)
\label{c10}
\end{eqnarray}
is the (reduced) collision frequency associated with the shear viscosity, where
\beq
{D}_{ij}(\mathbf{V})\equiv m\left(V_iV_j-\frac{V^2}{d}\delta_{ij}\right),
\label{n3.2}
\eeq
and
\begin{eqnarray}
\nu_\lambda^*&=&\frac{\int \dd\mathbf{v}\,\mathbf{S}\cdot{\cal L}^{(\alpha)}f_0\mathbf{S}}
{\nu_0\int \dd\mathbf{v}\,f_0\mathbf{S}\cdot\mathbf{S}}\nonumber\\
&=&
\frac{1+\alpha}{d}\left[\frac{d-1}{2}+\frac{3}{16}(d+8)(1-\alpha)\right.
\nonumber\\
&&\left.+
\frac{4+5d-3(4-d)\alpha}{512}a_2^\hcs\right]
\label{c11}
\end{eqnarray}
is the (reduced) collision frequency associated with the thermal conductivity, where
\beq
\mathbf{S}(\mathbf{V})\equiv \left(\frac{m}{2}V^2-\frac{d+2}{2}T\right)\mathbf{V}.
\label{n3.3}
\eeq
In the first equalities of Eqs.\ (\ref{c10}) and (\ref{c11}), ${\cal
L}^{(\alpha)}$ represents the linearization of the collision
operator $J^{(\alpha)}$ around the homogeneous cooling state:
\beq
{\cal L}^{(\alpha)}\phi\equiv
-J^{(\alpha)}[\phi,f_\hcs]-J^{(\alpha)}[f_\hcs,\phi].
\label{n3.1}
\eeq

In the model (\ref{b4}) the transport coefficients are formally
given by Eqs.\ (\ref{c5})--(\ref{c7}), except that $a_2^\hcs\to 0$
(since the homogeneous cooling state solution is now the local
equilibrium distribution, i.e., $f_\hcs=f_0$) and $\nu_\eta^*$ and
$\nu_\lambda^*$ are given by the first equalities of Eqs.\
(\ref{c10}) and (\ref{c11}) with the replacement
\begin{equation}
{\cal L}^{(\alpha)}\to \beta {\cal L}^{(1)}-\frac{\zeta_0}{2}\frac{\partial}{\partial\mathbf{v}}\cdot \mathbf{V},
\label{c11.2}
\end{equation}
where the operator $\mathcal{L}^{(1)}$ is defined by Eq.\ (\ref{n3.1}) with $\alpha=1$ and $f_\hcs=f_0$.
Taking into account the properties
\begin{equation}
\mathbf{V}\cdot \partial_\mathbf{v} D_{ij}=2D_{ij},\quad \mathbf{V}\cdot \partial_\mathbf{v} S_{i}=3S_{i}+(d+2)T V_i,
\label{c11.3}
\end{equation}
one easily gets
\begin{equation}
\nu_\eta^*\to \beta+\zeta_0^*,
\label{c12}
\end{equation}
\begin{equation}
\nu_\lambda^*\to \frac{d-1}{d}\beta+\frac{3}{2}\zeta_0^*.
\label{c13bis}
\end{equation}
In summary, the transport coefficients of the EHS model in the first
Sonine approximation are
\begin{equation}
\eta=\frac{nT}{\nu_0}\frac{1}{\beta+\frac{1}{2}\zeta_0^*},
\label{c13}
\end{equation}
\begin{equation}
\lambda=\frac{nT}{m\nu_0}\frac{d+2}{2}\frac{1}{\frac{d-1}{d}\beta-\frac{1}{2}\zeta_0^*},
\label{c14}
\end{equation}
\begin{equation}
\mu=\frac{T^2}{m\nu_0}\frac{d(d+2)}{2(d-1)}\frac{\zeta_0^*}{\beta(\frac{d-1}{d}\beta-\frac{1}{2}\zeta_0^*)}.
\label{c15}
\end{equation}

So far, the choice of the parameter $\beta$ remains open. To
reproduce the main trends in the $\alpha$-dependence of the shear
viscosity for IHS, let us equate Eq.\ (\ref{c13}) to Eq.\ (\ref{c5})
(with $a_2^\hcs=0$ for consistency). This yields
\begin{equation}
\beta=\frac{1+\alpha}{2}\left[1-\frac{d-1}{2d}(1-\alpha)\right]\equiv \beta_\eta.
\label{c16}
\end{equation}
Analogously, the model captures the behavior of the thermal
conductivity if $\beta$ is obtained by equating Eq.\ (\ref{c14}) to
Eq.\ (\ref{c6}) with $a_2^\hcs=0$. The result is
\begin{equation}
\beta=\frac{1+\alpha}{2}\left[1+\frac{3}{8}\frac{4-d}{d-1}(1-\alpha)\right]\equiv \beta_\lambda.
\label{c17}
\end{equation}
This expression also optimizes the agreement between Eqs.\ (\ref{c7}) and (\ref{c15}), i.e., $\beta_\mu=\beta_\lambda$.
\subsection{Self-diffusion}
If in the homogeneous cooling state a group of tagged particles
(here labeled with the subscript $\tg$) have initially a nonuniform
density $n_\tg$, the associated velocity distribution function
$f_\tg$ obeys the Boltzmann--Lorentz equation
\beq
\left(\partial_t+\mathbf{v}\cdot\nabla\right)f_\tg=J^{(\alpha)}[f_\tg,f_{\hcs}]\equiv -\mathcal{L}^{(\alpha)}_{\text{BL}}f_\tg.
\label{n3.5}
\eeq
As a consequence, a current
$\mathbf{j}_\tg=\int\dd\mathbf{v}\,\mathbf{v}f_\tg(\mathbf{v})$ of
tagged particles appears opposing the concentration gradient.
Conservation of the number of tagged particles implies the
continuity equation
\beq
\partial_t n_\tg+\nabla\cdot \mathbf{j}_\tg=0.
\label{n3.13}
\eeq
 In the limit of weak gradients,
one has
\beq
\mathbf{j}_\tg=-D\nabla n_\tg,
\label{n3.4}
\eeq
where $D$ is the self-diffusion coefficient. By standard application
of the Chapman--Enskog method in the first Sonine approximation, the
self-diffusion coefficient of IHS can be derived. The result is
\cite{BRCG00}
\beq
D=\frac{T}{m\nu_0}\frac{1}{\nu_D^*-\frac{1}{2}\zeta^*},
\label{n3.6}
\eeq
where
\begin{eqnarray}
\nu_D^*&=&\frac{\int \dd\mathbf{v}\,\mathbf{v}\cdot{\cal L}_{\text{BL}}^{(\alpha)}f_0\mathbf{v}}{\nu_0\int \dd\mathbf{v}\,v^2f_0}\nonumber\\
&=&\frac{d+2}{4d}(1+\alpha)\left(1
-\frac{1}{32}a_2^\hcs\right)
\label{n3.7}
\end{eqnarray}
is the (reduced) collision frequency associated with the self-diffusion coefficient.

In our EHS model, the Boltzmann--Lorentz equation (\ref{n3.5}) is replaced by
\beqa
\left[\partial_t+\mathbf{v}\cdot\nabla-\frac{\zeta_0}{2}\frac{\partial}{\partial
\mathbf{v}}\cdot \left(\mathbf{v}-\mathbf{u}_\tg\right) \right]f_\tg&=&\beta J^{(1)}[f_\tg,f_{0}]\nn
&\equiv& -\beta\mathcal{L}^{(1)}_{\text{BL}}f_\tg,
\label{n3.5bis}
\eeqa
where we have taken into account that the peculiar velocity
$\mathbf{V}$ appearing on the left-hand side of Eq.\ (\ref{b4}) must
now be understood as $\mathbf{v}-\mathbf{u}_\tg$, where
$\mathbf{u}_\tg=\mathbf{j}_\tg/n_\tg$ is the mean velocity of the
tagged particles, in order to make Eq.\ (\ref{n3.5bis}) consistent
with the continuity equation (\ref{n3.13}). Similarly to Eq.\
(\ref{c11.2}), the model implies
\begin{equation}
{\cal L}_{\text{BL}}^{(\alpha)}\to \beta {\cal L}_{\text{BL}}^{(1)}-\frac{\zeta_0}{2}\frac{\partial}{\partial\mathbf{v}}\cdot \mathbf{v},
\label{n3.9}
\end{equation}
so that
\beq
\nu^*_D\to \beta \frac{d+2}{2d}+\frac{\zeta_0^*}{2}.
\label{n3.10}
\eeq
In addition, the presence of the term proportional to $\mathbf{u}_t$
in Eq.\ (\ref{n3.5bis}) generates an extra term in the
Chapman--Enskog method such that  $\zeta^*/2$ in the denominator of
Eq.\ (\ref{n3.6}) is replaced by $\zeta_0^*$. In summary, the
self-diffusion coefficient corresponding to the EHS model is (in the
first Sonine approximation)
\beq
D=\frac{T}{m\nu_0}\frac{1}{\frac{d+2}{2d}\beta-\frac{1}{2}\zeta_0^*}.
\label{n3.11}
\eeq
Identifying Eq.\ (\ref{n3.11}) with Eq.\ (\ref{n3.6}) (by setting $a_2^\hcs=0$ in the latter), one gets
\beq
\beta=\frac{1+\alpha}{2}\equiv\beta_D.
\label{n3.12}
\eeq

\subsection{Comparison between the IHS and EHS transport coefficients}
\begin{figure}
 \includegraphics[width=.90 \columnwidth]{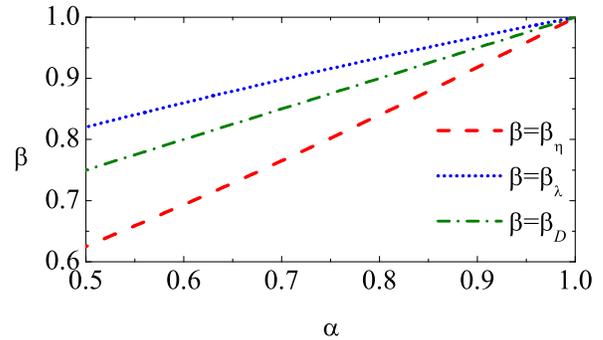}
 \caption{(Color online) Plot of $\beta_\eta$ (-- -- --), $\beta_\lambda$ ($\cdots$), and $\beta_D$
 (-- $\cdot$ -- $\cdot$ --) as functions of the coefficient of restitution $\alpha$ in the three-dimensional case.\label{beta}}
 \end{figure}
We have just seen that the suitable choice for the parameter
$\beta(\alpha)$ under the criterion of optimizing the agreement with
the transport coefficients of IHS is not unique. Depending on the
transport property of interest, it may be more convenient
$\beta=\beta_\eta$, $\beta=\beta_\lambda$, $\beta=\beta_D$, or even
a different choice. Focusing on these three possibilities, we note
that $\beta_\eta<\beta_D<\beta_\lambda$ for $d=2$ and $d=3$.
Moreover, $\beta_\lambda<1$ for all $\alpha$ in the case $d=3$, as
shown in Fig.\ \ref{beta}. This indicates that the diameter of the
equivalent EHS system is smaller than that of the actual IHS system
($\sigma'<\sigma$), this effect being more pronounced with
$\beta=\beta_\eta$ than with $\beta=\beta_\lambda$.

\begin{figure}
 \includegraphics[width=.90 \columnwidth]{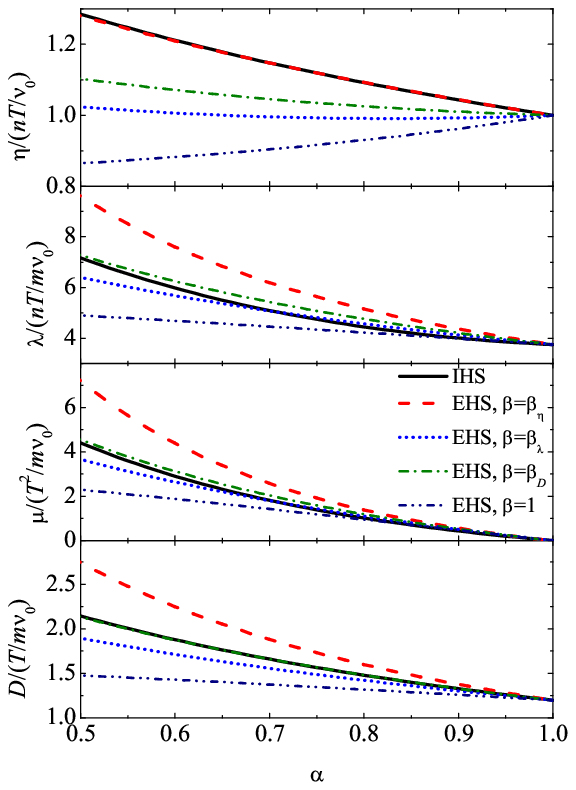}
 \caption{(Color online)  Plot of the (reduced) shear viscosity $\eta/(nT/\nu_0)$, thermal conductivity $\lambda/(nT/m\nu_0)$,
 transport coefficient $\mu/(T^2/m\nu_0)$, and self-diffusion coefficient $D/(T/m\nu_0)$
 for three-dimensional IHS (---) and the ``equivalent'' system of EHS
 with $\beta=\beta_\eta$ (-- -- --), $\beta=\beta_\lambda$ ($\cdots$),
 $\beta=\beta_D$ (-- $\cdot$ -- $\cdot$ --), and $\beta=1$ (-- $\cdot\cdot$ -- $\cdot\cdot$ --)
 as functions of the coefficient of restitution $\alpha$.
 Note that in the top panel the IHS curve and the EHS curve with $\beta=\beta_\eta$ are
practically indistinguishable. The same happens in the bottom panel
between the IHS curve and the EHS curve with $\beta=\beta_D$.
\label{coeff}}
 \end{figure}
Figure \ref{coeff}  compares the four transport coefficients of IHS
($d=3$) [Eqs.\ (\ref{c5})--(\ref{c7}) and (\ref{n3.6})] with  those
of the ``equivalent'' system of EHS [Eqs.\ (\ref{c13})--(\ref{c15})
and (\ref{n3.11})] with the choices $\beta=\beta_\eta$ [Eq.\
(\ref{c16})],  $\beta=\beta_\lambda$ [Eq.\ (\ref{c17})],
$\beta=\beta_D$ [Eq.\ (\ref{n3.12})],  and $\beta=1$. It can be
observed that the EHS shear viscosity  combined with the choice
$\beta=\beta_\eta$ reproduces almost perfectly the IHS shear
viscosity. A similar situation occurs in the case of the
self-diffusion coefficient with the choice $\beta=\beta_D$. This is
because the influence of $a_2^\hcs\neq 0$ in the cooling rate
$\zeta^*$ [Eq.\ (\ref{c9})] and in the collision frequencies
$\nu_\eta^*$ [Eq.\ (\ref{c10})] and $\nu_D^*$ [Eq.\ (\ref{n3.7})] is
very small. On the other hand, the deviations of the EHS
coefficients $\lambda$ and $\mu$ with $\beta=\beta_\lambda$ from the
respective IHS coefficients are much more important, essentially due
to the explicit dependence on $a_2^\hcs$ of the numerators on the
right-hand sides of Eqs.\ (\ref{c6}) and (\ref{c7}), since the
influence of $a_2^\hcs$ on $\nu_\lambda^*$ is again rather weak.

A natural question that arises is whether there exists a common
choice for $\beta(\alpha)$ that reproduces reasonably well the
$\alpha$-dependence of the four transport coefficients. Figure
\ref{coeff} shows that $\beta=\beta_\eta$, which is excellent in the
case of $\eta$, strongly overestimates $\lambda$, $\mu$, and $D$.
Moreover, $\beta=\beta_\lambda$ is not clearly superior to $\beta_D$
in the cases of $\lambda$ and $\mu$, whereas it is  poorer in
reproducing $\eta$ and $D$. Obviously, the naive choice $\beta=1$
does not capture well the $\alpha$-dependence of the transport
coefficients, especially in the case of the shear viscosity.
Therefore, we propose to take $\beta=\beta_D=\frac{1}{2}(1+\alpha)$.
With this choice, the frictional gas of EHS  has practically the
same self-diffusion coefficient as the true IHS gas and similar
values for the coefficients defining the heat flux. Although the
choice $\beta=\frac{1}{2}(1+\alpha)$ underestimates the shear
viscosity, this is not a serious drawback since this is the
transport coefficient least sensitive to inelasticity. For instance,
the ratio between the value of a transport coefficient at
$\alpha=0.5$ and the value at $\alpha=1$ is about 1.3 in the case of
$\eta$, while it is about 1.9 and 1.8 in the cases of $\lambda$ and
$D$, respectively. At that rather high inelasticity ($\alpha=0.5$),
the EHS transport coefficients with $\beta=\frac{1}{2}(1+\alpha)$
differ from the IHS ones by 14\% ($\eta$), 2\% ($\lambda$), 3\%
($\mu$), and 0.5\% ($D$).

There are two additional reasons to favor Eq.\ (\ref{n3.12}). First,
it is much simpler than Eqs.\ (\ref{c16}) and (\ref{c17}) and does
not depend on the dimensionality $d$. The second reason is more
compelling. The extension of the model (\ref{b4}) to dilute mixtures
and to dense gases is carried out in Appendices \ref{sec5.1} and
\ref{sec5.2}, respectively. In both cases the equation for the
collisional transfer of energy imposes the choice (\ref{n3.12}) as
the most natural one, without having to resort to the evaluation of
transport coefficients.

It is worth mentioning that the transport coefficients obtained from
frictional EHS with $\gamma=\frac{1}{2}\zeta_0$ and
$\beta=\frac{1}{2}(1+\alpha)$  exhibit a  much better  agreement
with the IHS coefficients than the ones  corresponding to the
inelastic Maxwell model \cite{S03}. The relationship between the
inelastic Maxwell model and the (frictional) elastic Maxwell model
is discussed in Appendix \ref{sec5.3}.

Before closing this Section, a comment is in order. The transport
coefficients $\eta$, $\lambda$, $\mu$, and $D$ have been considered
here in the first Sonine approximation, both for IHS and EHS, what
has allowed us to work with explicit expressions. Without
prejudicing the degree of reliability of the first Sonine
approximation, it can be understood as a useful tool to probe the
 structure of the linearized collision
operator through some relevant inner products [see the first
equalities of Eqs.\ (\ref{c10}), (\ref{c11}), and (\ref{n3.7})]. On
the other hand, a comparison of the first Sonine approximation for
the transport coefficients with direct simulation data shows a good
agreement for $\eta$ \cite{BR04,MSG04} and $D$ \cite{BRCG00,GM04},
but important discrepancies for high inelasticity are present in the
cases of $\lambda$ and $\mu$ \cite{BR04,MSG05}.

\section{Kinetic modeling\label{sec7}}
\subsection{BGK model}
The (approximate) mapping IHS$\leftrightarrow$EHS allows one to take
advantage of the existence of simple models for EHS to extend them
straightforwardly to IHS. For instance, consider the well-known
Bhatnagar--Gross--Krook (BGK) model for \textit{elastic} hard
spheres \cite{BGK54}:
\begin{equation}
J^{(1)}[f,f]\to -\nu_0(f-f_0),
\label{7.1}
\end{equation}
where $f_0$ is the local equilibrium distribution (\ref{3.5.1}) and
the effective collision frequency $\nu_0$ is usually identified with
Eq.\ (\ref{3.14b}) in order  to make the shear viscosity agree with
that of the Boltzmann equation. Thus, in consistency with the
approximation (\ref{b4}), the extension of the BGK model to
\textit{inelastic} hard spheres would simply be
\begin{equation}
J^{(\alpha)}[f,f]\to -\beta \nu_0(f-f_0)+\frac{\zeta_0 }{2}\frac{\partial }{\partial
{\bf v}}\cdot \left( {\bf V}f\right)
\label{7.2}
\end{equation}
with $\zeta_0$ and $\beta$ given by Eqs.\ (\ref{3.14}) and (\ref{n3.12}), respectively.
In fact, the kinetic model (\ref{7.2}) can be seen as a simplification of the one already proposed in Ref.\ \cite{BDS99}:
\beq
J^{(\alpha)}[f,f]\to -\beta \nu_0(f-f_\hcs)+\frac{\zeta }{2}\frac{\partial }{\partial
{\bf v}}\cdot \left( {\bf V}f\right),
\label{7.2bis}
\end{equation}
where here $f_\hcs$ represents the \textit{local} form of the
homogeneous cooling state, $\zeta$ is given by Eq.\ (\ref{c9}), and
$\beta$ is assumed to be given by Eq.\ (\ref{c16}).

The linearized collision operator corresponding to the BGK model (\ref{7.2}) is simply
\begin{equation}
{\cal L}^{(\alpha)}\to\beta\nu_0-\frac{\zeta_0 }{2}\frac{\partial }{\partial
{\bf v}}\cdot {\bf V}.
\label{7.3}
\end{equation}
As a consequence, the relation (\ref{c12}) holds again but one has
$\nu_\lambda^*\to \beta+\frac{3}{2}\zeta_0^*$ instead of
(\ref{c13bis}). Therefore, the shear viscosity is given by Eq.\
(\ref{c13}), while the thermal conductivity and the coefficient
$\mu$ are
\begin{equation}
\lambda=\frac{nT}{m\nu_0}\frac{d+2}{2}\frac{1}{\beta-\frac{1}{2}\zeta_0^*},
\label{7.5}
\end{equation}
\begin{equation}
\mu=\frac{T^2}{m\nu_0}\frac{d+2}{2}\frac{\zeta_0^*}{\beta(\beta-\frac{1}{2}\zeta_0^*)},
\label{7.6}
\end{equation}
which differ from  Eqs.\ (\ref{c14}) and (\ref{c15}), respectively, by a factor  $(d-1)/d$ in front of $\beta$.

In the case of self-diffusion in the homogeneous cooling state, the BGK kinetic equation for tagged particles is
\beq
\left[\partial_t+\mathbf{v}\cdot\nabla-\frac{\zeta_0}{2}\frac{\partial}{\partial
\mathbf{v}}\cdot \left(\mathbf{v}-\mathbf{u}_\tg\right)
\right]f_\tg=-\beta \nu_0\left(f_\tg-\frac{n_\tg}{n}f_0\right),
\label{n7.1}
\eeq
so the Boltzmann--Lorentz operator becomes
\begin{equation}
{\cal L}_{\text{BL}}^{(\alpha)}\to \beta \nu_0-\frac{\zeta_0}{2}\frac{\partial}{\partial\mathbf{v}}\cdot \mathbf{v}
\label{n7.1bis}
\end{equation}
and one has $\nu_D^*\to\beta +\frac{3}{2}\zeta_0^*$ instead of (\ref{n3.10}). Thus,
\beq
D=\frac{T}{m\nu_0}\frac{1}{\beta-\frac{1}{2}\zeta_0^*},
\label{n7.2}
\eeq
which differs from Eq.\ (\ref{n3.11}).

The inability of the BGK model to reproduce simultaneously the
different transport coefficients is already present in the elastic
case and is the price to be paid by the inclusion of a single
collision frequency $\nu_0$. In particular, the BGK model yields the
value $\text{Pr}=1$ for the Prandtl number $\text{Pr}\equiv
(d+2)\eta/2m\lambda$ in the elastic limit, while the correct
Boltzmann value is (in the first Sonine approximation)
$\text{Pr}=(d-1)/d$.

\subsection{Ellipsoidal statistical  model}
To avoid in part the above limitation, the so-called ellipsoidal
statistical (ES) model was proposed about forty years ago for
elastic particles \cite{C88,H66}. This kinetic model reads
\begin{equation}
J^{(1)}[f,f]\to -\nu_0\text{Pr}(f-f_R),
\label{7.8}
\end{equation}
where
\begin{equation}
f_R(\mathbf{v})=n\left(\frac{mn}{2\pi}\right)^{d/2} \left(\det
\mathsf{R}\right)^{-1/2}\exp\left(-\frac{mn}{2}\mathsf{R}^{-1}:\mathbf{V}\mathbf{V}\right)
\label{6.5:n2}
\end{equation}
is an anisotropic Gaussian distribution with the tensor $\mathsf{R}$ given by
\begin{equation}
\mathsf{R}=\frac{1}{\text{Pr}}\left[p\mathsf{I}-\left(1-\text{Pr}\right)\mathsf{P}\right],
\label{6.5:n3}
\end{equation}
$\mathsf{P}$ being the pressure tensor. The ES choice of $f_R$ is
based on information theory arguments. Note that the ES model
reduces to the conventional BGK model in the special case
$\text{Pr}=1$. It is then convenient to consider $\Pra$ as a free
parameter of the model so that the BGK model is recovered by
formally setting $\text{Pr}=1$. The reference distribution $f_R$ has
a finite norm provided that $\mathsf{R}$ is a positive definite
matrix, i.e., its eigenvalues $r_i$ must be non-negative. From Eq.\
(\ref{6.5:n3}), $r_i=\text{Pr}^{-1}[p-(1-\text{Pr})p_i]$, where
$p_i$ are the eigenvalues of the pressure tensor $\mathsf{P}$. Since
$\sum_{i=1}^d p_i=dp$, then $p_i\leq dp$ and, consequently, the
positiveness of $r_i$ implies that $\text{Pr}\geq ({d-1})/d$. The
lower bound coincides with the physical value of the  Prandtl
number. The first few moments of $f_R$ are
\begin{equation}
\int \dd\mathbf{v}\, \{1, \mathbf{V},m\mathbf{V}\mathbf{V}\}f_R(\mathbf{v})=\{n,\mathbf{0},\mathsf{R}\}.
\label{1.5:19}
\end{equation}
While in the BGK equation (\ref{7.1}) the reference function $f_0$
(namely, the local equilibrium distribution) is a functional of $f$
through its  hydrodynamic fields $n$, $\mathbf{u}$, and $T$, in the
ES model $f_R$ depends also on the irreversible part of the momentum
flux.

When (\ref{7.8}) is inserted into (\ref{b4}) we get our extension of
the ES kinetic model for IHS:
\begin{equation}
J^{(\alpha)}[f,f]\to -\beta\nu_0\text{Pr}(f-f_R)+\frac{\zeta_0 }{2}\frac{\partial }{\partial
{\bf v}}\cdot \left( {\bf V}f\right).
\label{7.9}
\end{equation}
We note that the ES model (\ref{7.9}) is different from the more
detailed Gaussian kinetic model recently proposed by Dufty et al.\
\cite{DBZ04}, the most important difference being that the collision
frequency becomes a function of the peculiar velocity in the latter
Gaussian model. The kinetic model (\ref{7.9}) should also be
distinguished from the ansatz of an anisotropic Gaussian (or
maximum-entropy) velocity distribution function introduced by
Jenkins and Richman \cite{JR88} as a means to obtain a closed set of
equations for the pressure tensor.

The linearized collision operator associated with (\ref{7.9}) is
given by (\ref{c11.2}), where now  the action of the linearized
operator in the elastic case is \cite{GS03}
\begin{eqnarray}
\mathcal{L}^{(1)}\phi(\mathbf{v})&=&\nu_0\left[\text{Pr}\phi(\mathbf{v})+\frac{1-\text{Pr}}{2pT}f_0(\mathbf{V})\mathsf{D}(\mathbf{V})\right.
\nonumber\\
&&\left.:\int \dd\mathbf{v}'\,\mathsf{D}\left(\mathbf{V}'\right)\phi(\mathbf{v}')\right].
\label{7.11}
\end{eqnarray}
This implies that the shear viscosity is given by Eq.\ (\ref{c13}),
while the thermal conductivity and the $\mu$ coefficient are
\begin{equation}
\lambda=\frac{nT}{m\nu_0}\frac{d+2}{2}\frac{1}{\text{Pr}\beta-\frac{1}{2}\zeta_0^*},
\label{7.13}
\end{equation}
\begin{equation}
\mu=\frac{T^2}{m\nu_0}\frac{d+2}{2}\frac{\zeta_0^*}{\text{Pr}\beta(\text{Pr}\beta-\frac{1}{2}\zeta_0^*)},
\label{7.14}
\end{equation}
respectively. Equations (\ref{7.13}) and (\ref{7.14}) coincide with
Eqs.\ (\ref{c14}) and (\ref{c15}) if one sets $\text{Pr}=(d-1)/d$.
Of course,  the BGK results (\ref{7.5}) and (\ref{7.6}) are
recovered by formally setting $\Pra=1$.

The natural version of the ES model to the self-diffusion problem is
\beq
\left[\partial_t+\mathbf{v}\cdot\nabla-\frac{\zeta_0}{2}\frac{\partial}{\partial
\mathbf{v}}\cdot \left(\mathbf{v}-\mathbf{u}_\tg\right) \right]f_\tg=-\beta \nu_0\Pra\left(f_\tg-\frac{n_\tg}{n}f_0\right).
\label{n7.3}
\eeq
Therefore,
\beq
D=\frac{T}{m\nu_0}\frac{1}{\Pra\beta-\frac{1}{2}\zeta_0^*},
\label{n7.2bis}
\eeq
which differs from Eq.\ (\ref{n3.11}), unless one allows the Prandtl number to take the artificial value $\Pra=(d+2)/2d$.

\subsection{Solution of the BGK and ES models for uniform shear flow\label{sec6}}
As an illustration of the BGK and ES models extended to IHS, let us
analyze their solutions in the case of one of the paradigmatic
nonequilibrium states, namely the uniform (or simple) shear flow. In
most of this Subsection we will consider the nonlinear ES model
(\ref{7.9}) with an arbitrary value for $\Pra$, so that $\Pra=1$
corresponds to the BGK model and $\Pra=(d-1)/d$ corresponds to the
true ES model.

In the uniform shear flow, the density is  constant, the granular
temperature is uniform, and the flow velocity has a linear profile
$\mathbf{u}=a y \widehat{\mathbf{x}}$, $a$ being the constant shear
rate. At a more fundamental level, the velocity distribution
function becomes uniform when the velocities are referred to the
co-moving Lagrangian frame:
\begin{equation}
f(\mathbf{r},\mathbf{v};t)=f(\mathbf{V},t), \quad \mathbf{V}=\mathbf{v}-a y \widehat{\mathbf{x}}.
\label{6.1}
\end{equation}
In that case, the ES model kinetic equation reads
\begin{equation}
\partial_t f-a V_y\frac{\partial}{\partial V_x}f-\frac{\zeta_0 }{2}\frac{\partial }{\partial
{\bf V}}\cdot \left( {\bf V}f\right)=-\beta\nu_0\text{Pr}(f-f_R).
\label{6.2}
\end{equation}
Multiplying both sides by $mV_iV_j$ and integrating over velocity, we get
\begin{equation}
\partial_t P_{ij}+a\left(\delta_{ix}P_{yj}+\delta_{jx}P_{iy}\right)+\zeta_0P_{ij}=-\beta\nu_0\left(P_{ij}-p\delta_{ij}\right),
\label{6.3}
\end{equation}
where on the right-hand side we have made use of Eqs.\
(\ref{6.5:n3}) and (\ref{1.5:19}). The set of equations (\ref{6.3})
is common to the BGK and the ES models since the constant $\Pra$
does not appear. The structure of Eq.\ (\ref{6.3}) is also obtained
from other BGK-like models \cite{BMD96,BRM97}, as well as from the
Boltzmann equation in Grad's approximation \cite{G02,SGD03}. In this
latter case, however, the flexibility of accommodating the
coefficient $\beta$ disappears since by construction it is
constrained to $\beta=\beta_\eta$.

The three independent equations stemming from Eq.\ (\ref{6.3}) are
\begin{equation}
\partial_t p+\zeta_0 p+\frac{2a}{d}P_{xy}=0,
\label{6.4}
\end{equation}
\begin{equation}
\partial_t P_{xy}+\left(\beta\nu_0+\zeta_0\right)P_{xy}+aP_{yy}=0,
\label{6.5}
\end{equation}
\begin{equation}
\partial_t P_{yy}+\left(\beta\nu_0+\zeta_0\right)P_{yy}-\beta\nu_0 p=0.
\label{6.6}
\end{equation}
Their steady-state solution is
\begin{equation}
{T}=\frac{T_0}{\nu_0^2(T_0)}\frac{2a^2}{d}\frac{\beta}{\zeta_0^*(\beta+\zeta_0^*)^2},
\label{6.7}
\end{equation}
\begin{equation}
\frac{P_{yy}}{nT}
=\frac{\beta}{\beta+\zeta_0^*},
\label{6.8}
\end{equation}
\begin{equation}
\frac{P_{xy}}{nT}
=- \sqrt{\frac{d}{2}}\frac{ \sqrt{\beta\zeta_0^*}}{\beta+\zeta_0^*},
\label{6.9}
\end{equation}
where in Eq.\ (\ref{6.7}) $T_0$ is an arbitrary reference
temperature and $\nu_0(T_0)$ is its associated collision frequency.
Except $P_{xy}=P_{yx}$, the remaining off-diagonal elements of the
pressure tensor vanish. In addition, $P_{yy}=P_{zz}=\cdots=P_{dd}$,
so that $P_{xx}=dp-(d-1)P_{yy}$. Equation (\ref{6.7}) can be
rewritten as
\beq
\frac{a}{\nu_0(T)}=\sqrt{\frac{d\zeta_0^*}{2\beta}}\left(\beta+\zeta_0^*\right),
\label{6.7bis}
\eeq
which gives the shear rate in units of the steady-state collision
frequency. The rheology of the uniform shear flow can be
conveniently characterized by a (dimensionless)
\textit{non-Newtonian} viscosity coefficient
\beq
\eta^*\equiv-\frac{P_{xy}}{nT}\frac{\nu_0(T)}{a}=\frac{\beta}{(\beta+\zeta_0)^2},
\label{n7.8}
\eeq
where use has been made of Eqs.\ (\ref{6.9}) and (\ref{6.7bis}).

\begin{figure}
 \includegraphics[width=.90 \columnwidth]{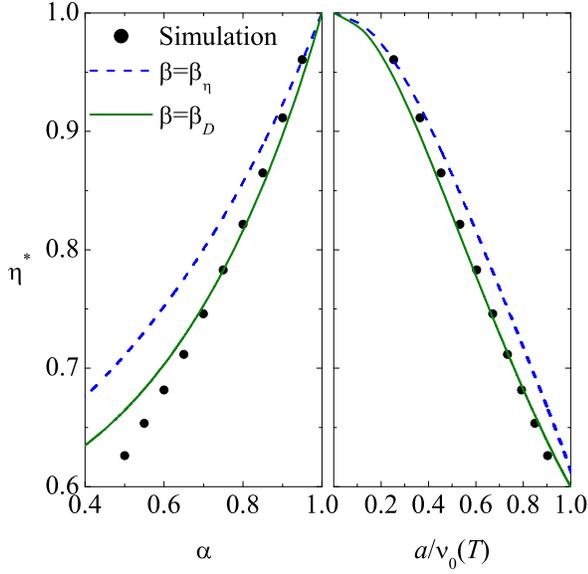}
 \caption{(Color online) Steady-state non-Newtonian shear viscosity for three-dimensional inelastic
hard spheres under uniform shear flow as a function of the
coefficient of restitution $\alpha$ (left panel) and of the reduced
shear rate $a/\nu_0$ (right panel). The curves are the common
predictions of the BGK and ES kinetic models, Eqs.\
(\protect\ref{6.7bis}) and (\protect\ref{n7.8}), with
$\beta=\beta_\eta$ (dashed lines) and $\beta=\beta_D=(1+\alpha)/2$
(solid lines). The circles are simulation results
\protect\cite{AS05}.\label{shear_BGK}}
 \end{figure}
Figure \ref{shear_BGK} shows the nonlinear shear viscosity
(\ref{n7.8}) as a function of the coefficient of restitution
$\alpha$ and of the reduced shear rate $a/\nu_0(T)$ for $d=3$ and
two choices of $\beta$: $\beta=\beta_\eta$ and
$\beta=\beta_D=\frac{1}{2}(1+\alpha)$. Numerical data obtained from
Monte Carlo simulations of the Boltzmann equation for IHS
\cite{AS04,AS05} are also shown. We observe that, except perhaps in
the quasi-elastic limit, the choice  $\beta=\beta_D$ exhibits a
better global agreement than the choice $\beta=\beta_\eta$, even
though the latter is tuned to reproduce the \textit{Newtonian} shear
viscosity (see Fig.\ \ref{coeff}). Since, as said above, Eqs.\
(\ref{6.3})--(\ref{n7.8}) with $\beta=\beta_\eta$ are derived from
the original Boltzmann equation for IHS in Grad's approximation, we
remark that the BGK and ES kinetic models with the choice
$\beta=\frac{1}{2}(1+\alpha)$ are more accurate than Grad's
approximation of the Boltzmann equation for this particular state.
This paradoxical result is partly due to the  inherently
non-Newtonian character of the steady uniform shear flow
\cite{SGD03}. It is interesting to note that for large inelasticity
the kinetic models tend to underestimate the shear thinning effect
of $\eta^*$ (see left panel of Fig.\ \ref{shear_BGK}). Since they
tend to underestimate the value of $a/\nu_0(T)$ as well (not shown),
it turns out that the agreement in the plot $\eta^*$ versus
$a/\nu_0(T)$ is fairly good even for large values of the reduced
shear rate (see right panel of Fig.\ \ref{shear_BGK}).

A practical advantage of kinetic models is the possibility of
obtaining explicitly the velocity distribution function. The
stationary solution to the kinetic equation (\ref{6.2}) can be
expressed as
\begin{eqnarray}
f(\mathbf{V})&=&\beta\nu_0\text{Pr}\Lambda^{-1}f_R(\mathbf{V})\nn
&=&
\beta\nu_0\text{Pr}\int_0^\infty \dd s\, \exp\left(-\Lambda s\right)f_R(\mathbf{V}),
\label{n7.4}
\end{eqnarray}
where the operator $\Lambda$ is
\beq
\Lambda=\beta\nu_0\text{Pr}-\frac{d}{2}\zeta_0-a V_y\frac{\partial}{\partial V_x}-
\frac{\zeta_0 }{2}{\bf V}\cdot\frac{\partial }{\partial
{\bf V}}.
\label{n7.5}
\eeq
The operators $V_y\partial/\partial V_x$ and $\mathbf{V}\cdot\partial/\partial\mathbf{V}$ commute. Therefore,
\begin{eqnarray}
f(\mathbf{V})
&=&
\beta\nu_0\text{Pr}\int_0^\infty \dd s\, \exp\left[-\left(\beta\nu_0\text{Pr}-\frac{d}{2}\zeta_0\right)s\right]\nn
&&\times f_R \left(\e^{\zeta_0s/2}\left(\mathbf{V}+asV_y\widehat{\mathbf{x}}\right)\right),
\label{6.10}
\end{eqnarray}
where we have taken into account the properties
\beq
\exp\left({as V_y\frac{\partial}{\partial V_x}}\right)\phi(\mathbf{V})=\phi\left(\mathbf{V}+asV_y\widehat{\mathbf{x}}\right),
\label{n7.6}
\eeq
\beq
\exp\left({
\frac{\zeta_0 }{2}s{\bf V}\cdot\frac{\partial }{\partial
{\bf V}}}\right)\phi(\mathbf{V})=\phi\left(\e^{\zeta_0s/2}\mathbf{V}\right).
\label{n7.7}
\eeq

 \begin{figure}
 \includegraphics[width=.90 \columnwidth]{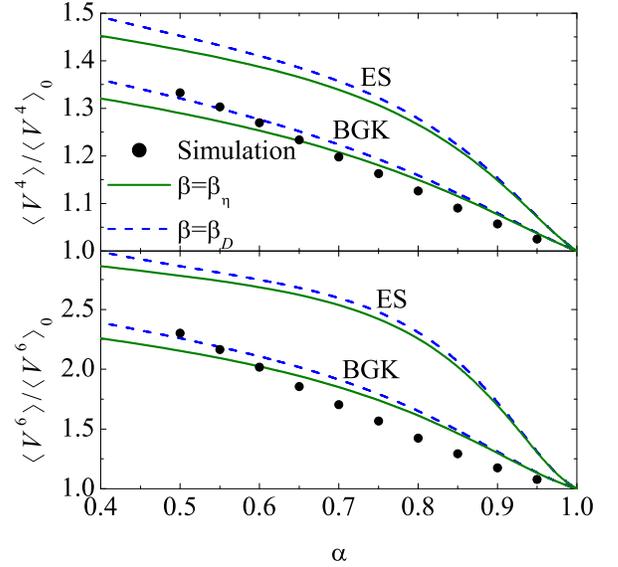}
\caption{(Color online) Steady-state fourth- and sixth-degree
moments, relative to their local equilibrium values, for
three-dimensional inelastic hard spheres under uniform shear flow as
functions of the coefficient of restitution $\alpha$. The curves are
the  predictions of the BGK and ES kinetic models with
$\beta=\beta_\eta$ (dashed lines) and $\beta=\beta_D=(1+\alpha)/2$
(solid lines). The circles are simulation results
\protect\cite{AS05}.\label{moment_BGK}}
 \end{figure}
Since the pressure tensor, and hence the reference distribution
function $f_R$, are entirely known, Eq.\ (\ref{6.10}) gives the
explicit solution. {}From it one can compute any desired velocity
moment. Some of those moments are derived in  Appendix \ref{appA}.
Figure \ref{moment_BGK} shows $\langle V^4\rangle/\langle
V^4\rangle_0$ and $\langle V^6\rangle/\langle V^6\rangle_0$, where
$\langle V^{2k}\rangle_0=(2T/m)^k \Gamma(k+d/2)/\Gamma(d/2)$ is the
local equilibrium value, as given by the BGK ($\Pra=1$) and ES
($\Pra=(d-1)/d)$) kinetic models, as well as by Monte Carlo
simulations of the Boltzmann equation for IHS \cite{AS05}. Even
though the ES model is more sophisticated than the BGK model, it
gives much poorer predictions for the fourth- and sixth-degree
moments than the BGK model. This situation is similar to that found
in the elastic case \cite{G97}. While this appears as a paradoxical
result, one must bear in mind that the real advantage of the ES
model over the BGK model occurs when the Prandtl number plays a
role, i.e., in states where momentum and energy transfer coexist.
Since in the steady uniform shear flow no energy transport is
present, there is no  reason \textit{a priori} to expect the ES
model to perform better than the BGK model. On the other hand, in
the steady Couette flow (where a quasi-parabolic temperature profile
coexists with a quasi-linear velocity profile), both models differ
already at the level of the non-Newtonian transport properties, the
ES model presenting in general a better agreement with simulation
results \cite{GS03,GH97}. In what concerns the influence of $\beta$,
we observe in Fig.\ \ref{moment_BGK} that
$\beta=\beta_D=(1+\alpha)/2$ is better for small or moderate
inelasticity, while $\beta=\beta_\eta$ tends to be better for large
inelasticity, an effect that contrasts with that of Fig.\
\ref{shear_BGK}.

\section{Concluding remarks\label{sec8}}
When one aims at understanding the basic properties of fluidized
granular media, the prototypical model consists of a system of
inelastic hard spheres (IHS) with a constant coefficient of
restitution $\alpha$. At a microscopic level of description, kinetic
theory has proven to be an efficient tool to unveil some of the most
intriguing features of IHS, such as the lack of equilibrium Gibbsian
states, high energy overpopulations, clustering instabilities,
Maxwell-demon effects, breakdown of energy equipartition, heat
fluxes induced by  density gradients, segregation in mixtures, phase
transitions in velocity space, \ldots The kinetic theory description
is based on the Boltzmann and Enskog equations for dilute and dense
granular gases, respectively. Both of them assume the molecular
chaos hypothesis, according to which the velocities of two particles
that are about to collide are uncorrelated. Although the  degree of
validity of this hypothesis is much more restricted in inelastic
collisions than in the case of elastic collisions \cite{SPM01}, the
Boltzmann and Enskog equations provide useful insights into the
peculiar behavior of granular fluids.

Mathematically speaking, the Boltzmann equation for IHS is much more
involved than for elastic hard spheres (EHS). As Eqs.\ (\ref{3.3})
and (\ref{3.2}) show, the coefficient of restitution $\alpha$
appears inside the velocity integral of the Boltzmann collision
operator in a two-fold way. First, it appears explicitly as a factor
$\alpha^{-2}$ in front of the gain term as a consequence of the
properties
$\dd\mathbf{v}''\dd\mathbf{v}_1''=\alpha^{-1}\dd\mathbf{v}\dd\mathbf{v}_1$
and ${\bf g}''\cdot \widehat{\bm{\sigma}}=-\alpha^{-1} {\bf g}\cdot
\widehat{\bm{\sigma}}$. Second, the coefficient of restitution
appears through the pre-collision velocities $\mathbf{v}''$ and
$\mathbf{v}_1''$ given by the collision rule (\ref{3.3}). The
primary consequence of $\alpha<1$ is the collisional loss of energy,
Eqs.\ (\ref{n1.1}) and (\ref{3.4}), which takes place at a cooling
rate $\zeta$ given by Eqs.\ (\ref{3.5}) and (\ref{3.5bis}).

The point we have addressed in this paper is the proposal of a model
of granular gases based on the Boltzmann equation for elastic hard
spheres (EHS). In this way, the complexities of inelastic collisions
for granular kinetic theory are represented by a simple two-fold
modification of the more familiar kinetic theory for elastic
collisions. Since elastic collisions conserve energy, the
fundamental ingredient of the model is the introduction of some kind
of external driving that produces a macroscopic cooling effect
similar to the one due to inelasticity. While the choice of that
driving is not unique, the most intuitive possibility consists of
assuming that the EHS gas is under the influence of a (fictitious)
drag force $\mathbf{F}_{\text{drag}}=-m\gamma\mathbf{V}$. In order
to produce the same cooling effect than in the true IHS gas, the
friction constant must be adjusted to be $\gamma=\frac{1}{2}\zeta$.
However, since in general $\zeta$ is a complicated nonlinear
functional of the one-particle velocity distribution function and we
want to keep the frictional EHS model as simple as possible, it is
preferable to take $\gamma=\frac{1}{2}\zeta_0$, where
$\zeta_0\propto n T^{1/2}(1-\alpha^2)$ is the cooling rate in the
local equilibrium approximation. The second ingredient of the model
is subtler. The application of the drag force is not enough for the
EHS gas to mimic, at least quantitatively, some other collisional
properties of the IHS gas, apart from the cooling effect. In our
model we have assumed that the collisions in the EHS gas are slowed
down by a factor $\beta<1$ with respect to the IHS gas. In the
dilute limit, this can be interpreted as assuming that the diameter
$\sigma'$ of EHS is smaller than the diameter $\sigma$ of  IHS,
namely $\sigma'/\sigma=\beta^{1/(d-1)}$. Alternatively, one can
interpret $\beta$  as a scaling factor in space and time. Comparison
between the Navier--Stokes transport coefficients derived from the
Boltzmann equation (in the first Sonine approximation) both for IHS
and frictional EHS suggests $\beta=\frac{1}{2}(1+\alpha)$. Moreover,
this choice is further supported by the mapping IHS$\to$EHS applied
to dilute gas mixtures (see Appendix \ref{sec5.1}) as well as to
dense single gases (see Appendix \ref{sec5.2}). In the latter case,
due to the physical separation between two colliding particles and
the presence of the pair correlation function in the Enskog
collision operator, the interpretation of $\beta$ in terms of
different sizes $\sigma'\neq\sigma$ or as a spatio-temporal scaling
factor is not literally correct. In that case, one must understand
$\beta$ just as a correction factor modifying the collision rate of
EHS relative to that of IHS.

It is obvious that an IHS gas and a gas made of frictional EHS
differ in many respects. For instance, the latter admits Maxwellians
as uniform solutions (with a time-dependent temperature in the case
of the homogeneous cooling state or as a stationary one in the case
of a white-noise forcing), while the IHS gas typically exhibits
overpopulated high energy tails. However, even though those
discrepancies might be qualitatively relevant, they are not
quantitatively  important in the domain of thermal speeds. In
particular, all but the two first distinctive features of granular
gases listed in the first paragraph of this Section can be expected
to appear in an EHS gas with a state-dependent friction constant
$\gamma\propto nT^{1/2}$.

When the gas is in anisotropic and/or inhomogeneous states  because
of either a transient stage or an instability or the application of
boundary conditions, transfers of momentum and/or energy are
present, so the velocity distribution function can deviate strongly
from a (local) Maxwellian, even for small and moderate velocities.
It is in those situations where the frictional EHS gas may be
expected to describe the main transport properties of the IHS gas.
We have tested this expectation by performing Monte Carlo
simulations of the Boltzmann equation under uniform shear flow both
for IHS and EHS \cite{AS04,AS05}. The results show that for moderate
inelasticity (say $\alpha=0.9$) one can hardly distinguish the
curves representing the transient profiles and the steady-state
velocity distribution functions corresponding to both systems. For
larger inelasticity (say $\alpha=0.5$) the EHS model still captures
almost entirely the nonequilibrium transport properties of the IHS
gas.

While the Boltzmann and Enskog equations for the EHS model are
mathematically more tractable than those for the true IHS system,
their solutions remain a formidable task. In the context of
conventional fluids, those difficulties have stimulated the proposal
of kinetic models, the prototype of which is the well-known BGK
model kinetic equation \cite{C88,BGK54}. An interesting practical
application of the relationship IHS$\leftrightarrow$EHS is the
possibility of extending those kinetic models to the case of
granular gases in a straightforward way. This has allowed us to
recover simplified versions of models previously proposed for
dilute  and dense single gases \cite{BDS99}. In the case of
inelastic mixtures, we are not aware of any previous proposal of a
kinetic model, despite the fact that several kinetic models exist
for elastic mixtures \cite{GS03}. Our method permits the
construction of extended kinetic models for inelastic mixtures, as
will be worked out elsewhere \cite{GSD04}.

\begin{acknowledgments}
It is a pleasure to thank V. Garz\'o and J. W. Dufty for a critical
reading of the manuscript. Partial support from the Ministerio de
Educaci\'on y Ciencia (Spain) through grant No.\ FIS2004-01399
(partially financed by FEDER funds) is gratefully acknowledged. A.A.
is grateful to the Fundaci\'on Ram\'on Areces (Spain) for a
predoctoral fellowship.
\end{acknowledgments}

\appendix
\section{Mixtures\label{sec5.1}}
\subsection{General properties}
In the case of a multi-component granular gas, the inelastic
collision between a sphere of species $i$ (mass $m_i$ and diameter
$\sigma_i$) and a sphere of species $j$ (mass $m_j$ and diameter
$\sigma_j$) is characterized by a coefficient of restitution
$\alpha_{ij}=\alpha_{ji}$. The direct and restituting collisions
rules are given by
\beq
\begin{array}{c}
{\bf v}'={\bf v}-\mu_{ji}\left(1+\alpha_{ij}\right)({\bf g}\cdot
\widehat{\bm{\sigma}})\widehat{\bm{\sigma}},\\
 {\bf v}_1'={\bf
v}_1+\mu_{ij}\left(1+\alpha_{ij}\right)({\bf g}\cdot
\widehat{\bm{\sigma}})\widehat{\bm{\sigma}},
\end{array}
\label{5.2bis}
\eeq
\beq
\begin{array}{c}
{\bf v}''={\bf v}-\mu_{ji}(1+\alpha_{ij}^{-1})({\bf
g}\cdot \widehat{\bm{\sigma}})\widehat{\bm{\sigma}}, \\
{\bf v}_1''={\bf v}_1+\mu_{ij}(1+\alpha_{ij}^{-1})({\bf g}\cdot
\widehat{\bm{\sigma}})\widehat{\bm{\sigma}},  \label{5.2}
\end{array}
\eeq
where
\begin{equation}
\mu_{ij}\equiv \frac{m_i}{m_i+m_j}.
\label{5.3}
\end{equation}
Equations (\ref{5.2bis}) and (\ref{5.2}) are generalizations of
Eqs.\ (\ref{3.3bis}) and (\ref{3.3}), respectively. Again, the
component of the post-collisional relative velocity along
$\widehat{\bm{\sigma}}$ is shrunk by a factor $\alpha_{ij}$, i.e.,
${\bf g}'\cdot \widehat{\bm{\sigma}}=-\alpha_{ij}{\bf g}\cdot
\widehat{\bm{\sigma}}$, and the kinetic energy decreases by a factor
proportional to $1-\alpha_{ij}^2$, namely
\beq
m_i{v'}^2+m_j{v_1'}^2-m_i v^2-m_j v_1^2=-({\bf g}\cdot
\widehat{\bm{\sigma}})^2(1-\alpha_{ij}^2)\frac{m_im_j}{m_i+m_j}.
\label{n5.17}
\eeq

{}From the velocity distribution function $f_i(\mathbf{v})$ of species $i$ one can define the number density
\beq
n_i=\int\dd \mathbf{v}\, f_i(\mathbf{v}),
\label{n5.3}
\eeq
the mass density $\rho_i=m_in_i$, and the average velocity
\beq
\mathbf{u}_i=\frac{1}{n_i}\int\dd \mathbf{v}\, \mathbf{v}f_i(\mathbf{v})
\label{n5.4bis}
\eeq
of species $i$. The associated global quantities are the total
number density $n=\sum_in_i$, the total mass density
$\rho=\sum_i\rho_i$, and the  (barycentric) flow velocity
\begin{equation}
\mathbf{u}=\frac{1}{\rho}\sum_i \rho_i \mathbf{u}_i.
\label{5.8}
\end{equation}
The granular temperature $T$ of the mixture is defined by
\beq
T=\frac{1}{n}\sum_i n_iT_i, \quad \frac{d}{2}n_i T_i=\frac{m_i}{2}\int\dd\mathbf{v}\,V^2f_i(\mathbf{v}),
\label{n5.5}
\eeq
where $\mathbf{V}=\mathbf{v}-\mathbf{u}$ is the peculiar velocity.
In Eq.\ (\ref{n5.5}) $T_i$ is the \textit{partial} granular
temperature associated with species $i$. In general, equipartition
of energy does not hold, even in homogeneous states, so $T_i\neq T$
\cite{GD99}.

In the dilute limit, the distribution functions $f_i(\mathbf{v})$  obey a set of coupled Boltzmann equations,
\beq
\left(\partial_t+\mathbf{v}\cdot\nabla\right)f_i=\sum_j J_{ij}^{(\alpha_{ij})}[f_i,f_j],
\label{n5.1}
\eeq
where the collision operator $J_{ij}^{(\alpha_{ij})}[f_i,f_j]$ is given by
\begin{eqnarray}
&&J_{ij}^{(\alpha_{ij})}[f_i,f_j]=\sigma_{ij} ^{d-1}\int \dd{\bf v}_{1}\int \dd\widehat{\bm{\sigma}}\,\Theta (
{\bf g}\cdot \widehat{\bm{\sigma}})({\bf g}\cdot \widehat{\bm{\sigma}})\nonumber\\
&&\times\left[ \alpha_{ij} ^{-2}f_i({\bf v}'')f_j({\bf v}
_{1}'')-f_i({\bf v})f_j({\bf v}_{1})\right] ,
\label{5.1}
\end{eqnarray}
where $\sigma_{ij}\equiv (\sigma_i+\sigma_j)/2$.
Every collision conserves the number of particles of each species,
\beq
\int\dd \mathbf{v}\,J_{ij}^{(\alpha_{ij})}[f_i,f_j]=0.
\label{n5.2}
\eeq
Moreover, the \textit{total} momentum is conserved as well, i.e.,
\beq
\sum_{i,j}m_i\int\dd \mathbf{v}\,\mathbf{v}J_{ij}^{(\alpha_{ij})}[f_i,f_j]=\mathbf{0}.
\label{n5.3bis}
\eeq
However, the total energy is not conserved:
\beq
\sum_{i,j}\frac{m_i}{2}\int\dd \mathbf{v}\,V^2J_{ij}^{(\alpha_{ij})}[f_i,f_j]=-\frac{d}{2}nT\zeta,
\label{n5.4}
\eeq
what defines the collision rate $\zeta$ of the mixture.

In general, given an arbitrary function $\phi(\mathbf{v})$, one can define its associated collision integral
\begin{eqnarray}
I_{ij}^{(\alpha_{ij})}[\phi]&\equiv&\int \dd\mathbf{v}\phi(\mathbf{v})J_{ij}^{(\alpha_{ij})}[f_i,f_j]\nn
&=&
\sigma_{ij} ^{d-1}\int \dd\mathbf{v}\int \dd{\bf v}_{1}\int \dd\widehat{\bm{\sigma}}\,\Theta (
{\bf g}\cdot \widehat{\bm{\sigma}})({\bf g}\cdot \widehat{\bm{\sigma}})\nonumber\\
&&\times f_i({\bf v})f_j({\bf v}_{1})\left[\phi(\mathbf{v}')-\phi(\mathbf{v})\right],
\label{5.4}
\end{eqnarray}
where in the last step we have performed a standard change of variables.
Thus, Eqs.\ (\ref{n5.2})--(\ref{n5.4}) can be rewritten as
\beq
I_{ij}^{(\alpha_{ij})}[1]=0,
\label{n5.6}
\eeq
\beq
\sum_{i,j}I_{ij}^{(\alpha_{ij})}[m_i\mathbf{v}]=\mathbf{0},
\label{n5.7}
\eeq
\beq
\zeta=-\frac{1}{dnT}\sum_{i,j}I_{ij}^{(\alpha_{ij})}[m_iV^2].
\label{n5.8}
\eeq

The collision integrals $I_{ij}^{(\alpha_{ij})}[m_i\mathbf{v}]$ and
$I_{ij}^{(\alpha_{ij})}[m_iV^2]$ cannot be evaluated exactly for
arbitrary distribution functions $f_i$ and $f_j$. This situation is
analogous to the one taking place with Eq.\ (\ref{3.5bis}) in the
single gas case. Again, a reasonable estimate can be expected if the
integrals are evaluated in the (multi-temperature) \textit{Gaussian
approximation}
\begin{equation}
f_i(\mathbf{v})\to f_{i,0}(\mathbf{v})=n_i\left(\frac{m_i}{2\pi T_i}\right)^{d/2}\exp\left(-m_iV^2/2T_i\right),
\label{5.7}
\end{equation}
where we have restricted ourselves to the case
$\mathbf{u}_i=\mathbf{u}$ in order to satisfy Eq.\ (\ref{n5.4bis}).
When the approximation (\ref{5.7}) is inserted into Eq.\ (\ref{5.4})
one gets  \cite{GD02,GM03}
\beq
I_{ij}^{(\alpha_{ij})}[m_i\mathbf{v}]\to 0,
\label{n5.9}
\eeq
\begin{eqnarray}
I_{ij}^{(\alpha_{ij})}[m_iV^2]&\to&(d+2)n_iT_i\nu_{ij}\frac{1+\alpha_{ij}}{2}
\nn&&\times
\left[\frac{m_i(T_j-T_i)}{m_jT_i+m_iT_j}
-\frac{1-\alpha_{ij}}{2}\right],
\label{5.11}
\end{eqnarray}
where
\beq
\nu_{ij}=\frac{4\Omega_d}{\sqrt{\pi}(d+2)}n_j\mu_{ji}^2\sigma_{ij}^{d-1}\left(\frac{2T_i}{m_i}\right)^{1/2}\left(1+\frac{m_iT_j}{m_jT_i}\right)^{3/2}
\label{5.14}
\eeq
is an effective collision frequency of a particle of species $i$
with particles of species $j$. In this Gaussian approximation, the
cooling rate defined by Eq.\ (\ref{n5.8}) becomes $\zeta\to\zeta_0$
with
\beq
\zeta_0=\frac{d+2}{4dnT}\sum_{i,j}
n_iT_i\nu_{ij}\left(1-\alpha_{ij}^2\right),
\label{5.12}
\eeq
where we have made use of the property $\rho_iT_i\nu_{ij}=\rho_jT_j\nu_{ji}$.
In the case of a single gas, Eqs.\ (\ref{5.14}) and (\ref{5.12}) reduce to Eqs.\ (\ref{3.14b}) and (\ref{3.14}), respectively.

\subsection{Model of frictional elastic hard spheres}
Once we have revised some of the basic properties of the collision
operator $J_{ij}^{(\alpha_{ij})}[f_i,f_j]$ for an IHS mixture, we
are in conditions of proposing a \textit{minimal} model for
frictional EHS. In agreement with the philosophy behind Eq.\
(\ref{b4}), we write
\begin{equation}
J_{ij}^{(\alpha_{ij})}[f_i,f_j]\to \beta_{ij} J_{ij}^{(1)}[f_i,f_j]+\frac{\zeta_{ij}}{2}\frac{\partial }{\partial
{\bf v}}\cdot \left[\left( {\bf v }-\mathbf{u}_{i}\right)f_i\right],
\label{5.5}
\end{equation}
where $\beta_{ij}$ and $\zeta_{ij}$ are to be determined by
optimizing the agreement between the relevant properties of the true
operator $J_{ij}^{(\alpha_{ij})}[f_i,f_j]$ and those of the
right-hand side of Eq.\ (\ref{5.5}). For an arbitrary function
$\phi(\mathbf{v})$, one has
\begin{equation}
I_{ij}^{(\alpha_{ij})}[\phi]\to\beta_{ij}I_{ij}^{(1)}[\phi]-\frac{\zeta_{ij}}{2}\int \dd\mathbf{v}
\left( {\bf v }-\mathbf{u}_{i}\right)\cdot \frac{\partial \phi(\mathbf{v})}{\partial
{\bf v}} f_i(\mathbf{v}).
\label{5.6}
\end{equation}
The above replacement satisfies Eq.\ (\ref{n5.6}) identically. In
addition, it is consistent with Eq.\ (\ref{n5.9}) in the Gaussian
approximation (\ref{5.7}). So far, $\beta_{ij}$ and $\zeta_{ij}$
remain arbitrary. Insertion of the Gaussian approximation
(\ref{5.7}) into Eq.\ (\ref{5.6}) with $\phi(\mathbf{v})=m_iV^2$
yields
\begin{eqnarray}
I_{ij}^{(\alpha_{ij})}[m_iV^2]&\to&\beta_{ij}(d+2)n_iT_i\nu_{ij}\frac{m_i(T_j-T_i)}{m_jT_i+m_iT_j}\nn
&&
-\zeta_{ij}dn_iT_i,
\label{5.11bis}
\end{eqnarray}
where again we have restricted ourselves to the case
$\mathbf{u}_i=\mathbf{u}$. Comparison between Eqs.\ (\ref{5.11}) and
(\ref{5.11bis}) suggests the choices
\begin{equation}
\beta_{ij}=\frac{1+\alpha_{ij}}{2},
\label{5.13}
\end{equation}
\beq
\zeta_{ij}=\frac{d+2}{4d}\nu_{ij}\left(1-\alpha_{ij}^2\right).
\label{n5.10}
\eeq
Of course, other combinations of $\beta_{ij}$ and $\zeta_{ij}$ are
in principle possible, but Eqs.\ (\ref{5.13}) and (\ref{n5.10})
represent the simplest choice in the absence of mutual diffusion
(i.e., with $\mathbf{u}_i=\mathbf{u}$). In the more general case
$\mathbf{u}_i\neq\mathbf{u}$, the expression for $\zeta_{ij}$ is
more complicated and will be reported elsewhere \cite{GSD04}.

Equation (\ref{5.11bis}) highlights that in general
$I^{(\alpha_{ij})}[m_iV^2]\neq 0$ because of two reasons. First, if
species $i$ and $j$ have different mean kinetic energies (i.e.,
$T_i\neq T_j$), then mutual collisions  tend to ``equilibrate'' both
partial temperatures. This \textit{equipartition} effect, which is
also present in the case of elastic collisions, is represented by
the term $\beta_{ij}I^{(1)}[m_iV^2]$. Moreover, even if $T_i= T_j$,
one has $I^{(\alpha_{ij})}[m_iV^2]\neq 0$ due to the inelasticity of
collisions, an effect that is accounted for  by the term
$-\zeta_{ij}dn_iT_i$. While the former term can be either positive
($T_i<T_j$) or negative ($T_i>T_j$), the latter term is  negative
definite. Thus, $\zeta_{ij}$ represents the cooling rate of species
$i$ due to collisions with particles of species $j$. Since the
relative decrease of energy after each $i$-$j$ collision is
proportional to $1-\alpha_{ij}^2$, it is quite natural that
$\zeta_{ij}\propto 1-\alpha_{ij}^2$. As in the single gas case, the
parameter $\beta_{ij}$ measures the rate of $i$-$j$ collisions of
EHS relative to that of IHS. It is rather reinforcing that the
choice (\ref{5.13}) is the natural extension of the choice
(\ref{n3.12}) that we adopted for a single gas. While in the latter
case we had to resort to the evaluation of the transport
coefficients, the choice (\ref{5.13}) arises in the case of mixtures
simply from the collisional energy transfer when $T_i\neq T_j$.

\subsection{Brownian limit}
Let us consider now the Brownian limit of a heavy impurity particle
(species 1) immersed in a bath of light particles (species 2). In
that case, the Boltzmann--Lorentz operator for inelastic collisions
becomes the Fokker--Planck operator \cite{BDS99}
\begin{equation}
J_{12}^{(\alpha_{12})}[f_1,f_2]=\frac{1+\alpha_{12}}{2}J_{12}^{(1)}[f_1,f_2]-\frac{1}{2}\zeta_{12}\frac{T_1}{m_1}\frac{\partial^2}{\partial v^2}f_1,
\label{5.18}
\end{equation}
where
\begin{equation}
J_{12}^{(1)}[f_1,f_2]=\gamma_{12} \frac{\partial}{\partial \mathbf{v}}\cdot \left(\mathbf{v}+\frac{T_2}{m_1}
\frac{\partial}{\partial \mathbf{v}}\right)f_1.
\label{5.15}
\end{equation}
In the above equations,
\begin{equation}
\gamma_{12}=\frac{2\Omega_d}{d\sqrt{\pi}}n_2\sigma_{12}^{(d-1}(m_2/m_1)^{1/2}\left(\frac{2T_2}{m_1}\right)^{1/2}
\label{5.16}
\end{equation}
is the friction constant associated with elastic collisions and
\begin{equation}
\zeta_{12}=\frac{\gamma_{12}}{2}\frac{T_2}{T_1}(1-\alpha_{12}^2)
\label{5.17}
\end{equation}
is the cooling rate of the Brownian particle due to inelastic
collisions with the bath particles. Equation (\ref{5.18}) is the
exact Fokker--Plank limit of the inelastic Boltzmann--Lorentz
operator. It is quite nice that Eq.\ (\ref{5.18}) supports the
choice (\ref{5.13}) as the collision rate factor. Moreover, Eq.\
(\ref{5.17}) for $\zeta_{12}$ agrees with the limit $m_1\gg m_2$ of
the proposal  (\ref{n5.10}). On the other hand, the exact
Fokker--Planck equation (\ref{5.18}) differs from the one obtained
from our model (\ref{5.5}). More specifically, the  latter results
from the replacement
\begin{equation}
-\frac{\zeta_{12}}{2}\frac{T_1}{m_1}\frac{\partial^2}{\partial v^2}f_1\to
\frac{\zeta_{12}}{2}\frac{\partial }{\partial
{\bf v}}\cdot \left[({\bf v }-{\bf u }_1)f_1\right].
\label{5.19}
\end{equation}
While on the right-hand side of  (\ref{5.19}) the role of mimicking
the collisional cooling experienced by the Brownian particle is
played by a deterministic frictional force, that role is played by a
stochastic force on the left-hand side. On the other hand, both
sides of (\ref{5.19}) agree in that they give vanishing
contributions to the mass and momentum balance equations and yield
the same contribution to the energy balance equation if
$\mathbf{u}_1=\mathbf{u}$. In addition, they coincide in the
Gaussian approximation (\ref{5.7}). Therefore, the model (\ref{5.5})
can be expected to behave reasonably well even in the extreme limit
of a Brownian particle.

\subsection{Kinetic modeling}
As discussed in Sec.\ \ref{sec7}, the mapping  (\ref{5.5}) allows one to transfer any given kinetic model
\begin{equation}
J_{ij}^{(1)}[f_i,f_j]\to K_{ij}^{(1)}
\label{5.22}
\end{equation}
for \textit{elastic} mixtures \cite{GS03} into an equivalent model for \textit{inelastic} mixtures:
\begin{equation}
J_{ij}^{(\alpha_{ij})}[f_i,f_j]\to K_{ij}^{(\alpha_{ij})}= \beta_{ij}K_{ij}^{(1)}
+\frac{\zeta_{ij}}{2}\frac{\partial }{\partial
{\bf v}}\cdot \left[\left( {\bf v }-\mathbf{u}_{i}\right)f_i\right].
\label{5.23}
\end{equation}
Thus, one can construct in a straightforward way the inelastic
versions of kinetic models such as Gross--Krook's \cite{GK56},
Garz\'o--Santos--Brey's \cite{GSB89}, or Andries--Aoki--Perthame's
\cite{AAP02}. More details will be reported elsewhere \cite{GSD04}.

\section{Dense granular gases\label{sec5.2}}
\subsection{The Enskog equation for inelastic hard spheres}
So far, we have restricted ourselves to low density granular gases
for which the Boltzmann description seems appropriate. At higher
densities the revised Enskog kinetic theory, suitably generalized to
inelastic collisions \cite{BDS97}, provides a basis for analysis of
granular flow. The Enskog kinetic equation reads
\begin{equation}
\left( \partial _{t}+{\bf v\cdot \nabla }\right) f=J_E^{(\alpha)}[f],
\label{n5.12}
\end{equation}
where
\begin{eqnarray}
J_E^{(\alpha)}[f]=\sigma ^{d-1}\int \dd{\bf v}_{1}\int \dd\widehat{\bm{\sigma}}\,\Theta (
{\bf g}\cdot \widehat{\bm{\sigma}})({\bf g}\cdot \widehat{\bm{\sigma}})\nonumber\\
\times\left[ \alpha ^{-2}f_2({\bf r},{\bf r}-{\bm{\sigma}},{\bf v}'',{\bf v}''_1)-f_2({\bf r},{\bf r}+{\bm{\sigma}},{\bf v},{\bf v}_1)\right] .
\label{n5.13}
\end{eqnarray}
In this equation,  $\bm{\sigma}\equiv\sigma\widehat{\bm{\sigma}}$ and
\beq
f_2(\mathbf{r},\mathbf{r}\pm{\bm{\sigma}},\mathbf{v},\mathbf{v}_1)=
\chi(\mathbf{r},\mathbf{r}\pm{\bm{\sigma}})
f(\mathbf{r},\mathbf{v})f(\mathbf{r}\pm{\bm{\sigma}},\mathbf{v}_1)
\label{n5.14}
\eeq
is the pre-collisional two-body distribution function in the
molecular chaos approximation,
$\chi(\mathbf{r},\mathbf{r}\pm{\bm{\sigma}})$ being the equilibrium
pair correlation function at contact as a functional of the density
field. We use the notation $J_E^{(\alpha)}[f]$ rather than
$J_E^{(\alpha)}[f,f]$ to remind that the nonlinear functional
dependence of the Enskog collision operator on the velocity
distribution function $f$ is higher than bilinear, due to the
presence of the correlation function $\chi$.

Taking velocity moments in (\ref{n5.12}) one obtains again the
hydrodynamic balance equations (\ref{n1.3})--(\ref{n1.5}), where the
number density, flow velocity, and granular temperature are defined
by Eqs.\ (\ref{b1})--(\ref{b3}), respectively. The pressure tensor
and the heat flux have both kinetic and collisional transfer
contributions, namely $\mathsf{P}=\mathsf{P}_k+\mathsf{P}_c$,
$\mathbf{q} =\mathbf{q}_k+\mathbf{q}_c$. The kinetic contributions
$\mathsf{P}_k$ and $\mathbf{q}_k$ are given by Eqs.\ (\ref{c1}) and
(\ref{c2}), while the collisional transfer parts are \cite{BDS97}
\beqa
\mathsf{P}_c^{(\alpha)}[f]&=&\frac{1+\alpha}{4}m\sigma^{d}\int\dd\mathbf{v}
\int \dd{\bf v}_{1}\int \dd\widehat{\bm{\sigma}}\,\Theta (
{\bf g}\cdot \widehat{\bm{\sigma}})({\bf g}\cdot \widehat{\bm{\sigma}})^2\nonumber\\
&&\times\widehat{\bm{\sigma}}\widehat{\bm{\sigma}}
\int_0^1\dd\lambda\, f_2\left({\bf r}-(1-\lambda){\bm{\sigma}},{\bf
r}+\lambda{\bm{\sigma}},{\bf v},{\bf v}_1\right),\nn
\label{n5.15}
\eeqa
\beqa
\mathbf{q}_c^{(\alpha)}[f]&=&\frac{1+\alpha}{4}m\sigma^{d}\int\dd\mathbf{v}
\int \dd{\bf v}_{1}\int \dd\widehat{\bm{\sigma}}\,\Theta (
{\bf g}\cdot \widehat{\bm{\sigma}})({\bf g}\cdot \widehat{\bm{\sigma}})^2\nonumber\\
&&\times(\mathbf{G}\cdot\widehat{\bm{\sigma}})\widehat{\bm{\sigma}}
\int_0^1\dd\lambda\, f_2({\bf r}-(1-\lambda){\bm{\sigma}},{\bf
r}+\lambda{\bm{\sigma}},{\bf v},{\bf v}_1),\nn
\label{n5.16}
\eeqa
where $\mathbf{G}\equiv
(\mathbf{V}+\mathbf{V}_1)/2=(\mathbf{v}+\mathbf{v}_1)/2-\mathbf{u}(\mathbf{r})$.
The superscript $(\alpha)$ on $\mathsf{P}_c^{(\alpha)}$ and
$\mathbf{q}_c^{(\alpha)}$ has been introduced to emphasize that both
quantities depend \textit{explicitly} on the coefficient of
restitution. They also depend implicitly on $\alpha$ through their
functional dependence on the velocity distribution function $f$.
With that convention, we can write
\beq
\mathsf{P}_c^{(\alpha)}[f]=\frac{1+\alpha}{2}\mathsf{P}_c^{(1)}[f],\quad
\mathbf{q}_c^{(\alpha)}[f]=\frac{1+\alpha}{2}\mathbf{q}_c^{(1)}[f].
\label{n6.22}
\eeq
The cooling rate is found to be \cite{BDS97,BDS99}
\beqa
\zeta
&=&(1-\alpha^2)\frac{m\sigma^{d-1}}{4dnT}
\int\dd\mathbf{v}
\int \dd{\bf v}_{1}\int \dd\widehat{\bm{\sigma}}\,\Theta (
{\bf g}\cdot \widehat{\bm{\sigma}})({\bf g}\cdot \widehat{\bm{\sigma}})^3\nn
&&\times f_2({\bf r},{\bf r}+{\bm{\sigma}},{\bf v},{\bf v}_1).
\label{n6.16}
\eeqa
The divergences of the collisional transfer parts are simply related
to moments of the collision operator \cite{BDS97},
\beq
\nabla\cdot\mathsf{P}_c^{(\alpha)}=-m\int\dd\mathbf{V}\,\mathbf{V}J^{(\alpha)}_E[f],
\label{n6.18}
\eeq
\beq
\nabla\cdot\mathbf{q}_c^{(\alpha)}=-\frac{m}{2}\int\dd\mathbf{V}\,{V^2}J^{(\alpha)}_E[f]-\mathsf{P}_c^{(\alpha)}:\nabla\mathbf{u}-\frac{d}{2}n\zeta T.
\label{n6.19}
\eeq

As in the dilute case, a reasonable estimate of the cooling rate
$\zeta(\mathbf{r},t)$ can be expected if the velocity distribution
function is replaced by its local equilibrium approximation
(\ref{3.5.1}). The resulting cooling rate is
\beqa
\zeta_0(\mathbf{r})&=&(1-\alpha^2)\frac{m\sigma^{d-1}}{4\pi^ddT(\mathbf{r})}\int\dd\mathbf{C}\int\dd\mathbf{C}_1 \e^{-C^2-C_1^2}\nn
&&\times\int \dd\widehat{\bm{\sigma}}\,\Theta (
{\bf g}\cdot \widehat{\bm{\sigma}})({\bf g}\cdot \widehat{\bm{\sigma}})^3\chi(\mathbf{r},\mathbf{r}+\bm{\sigma})n(\mathbf{r}+\bm{\sigma}),\nn
\label{n6.17}
\eeqa
where now
$\mathbf{g}=\sqrt{2T(\mathbf{r})/m}\mathbf{C}-\sqrt{2T(\mathbf{r}+\bm{\sigma})/m}\mathbf{C}_1+\mathbf{u}(\mathbf{r})-\mathbf{u}(\mathbf{r}+\bm{\sigma})$.
The local equilibrium cooling rate $\zeta_0(\mathbf{r})$ depends not
only on the local values of the hydrodynamic fields at $\mathbf{r}$
but also on their values on a spherical surface of radius $\sigma$
around the point $\mathbf{r}$. It also depends on the entire density
field through $\chi(\mathbf{r},\mathbf{r}+\bm{\sigma})$. If one
further neglects those dependencies, one gets the same result as in
the dilute limit, Eqs.\ (\ref{3.14}) and (\ref{3.14b}), except that
the collision frequency $\nu_0$ is multiplied by $\chi$.

\subsection{Model of frictional elastic hard spheres}
The natural extension to the Enskog equation of the model (\ref{b4}) is
\begin{equation}
J_E^{(\alpha)}[f]\to \beta J^{(1)}_E[f]+\frac{\zeta_0 }{2}\frac{\partial }{\partial
{\bf v}}\cdot \left( {\bf V}f\right).
\label{n6.20}
\end{equation}
Insertion of (\ref{n6.20}) into Eqs.\ (\ref{n6.18}) and (\ref{n6.19}) yields
\beq
\nabla\cdot\mathsf{P}_c^{(\alpha)}\to \beta \nabla\cdot\mathsf{P}_c^{(1)},
\label{n6.21}
\eeq
\beq
\nabla\cdot\mathbf{q}_c^{(\alpha)}+\mathsf{P}_c^{(\alpha)}:\nabla\mathbf{u}\to
\beta
\left[\nabla\cdot\mathbf{q}_c^{(1)}+\mathsf{P}_c^{(1)}:\nabla\mathbf{u}\right],
\label{n6.23}
\eeq
where in Eq.\ (\ref{n6.23}) we have approximated $\zeta\to\zeta_0$.
Comparison with the exact results (\ref{n6.22}) implies that
$\beta=(1+\alpha)/2$. Again, this reinforces the choice
(\ref{n3.12}) made in the dilute case.

\subsection{Kinetic modeling}
A few years ago, Dufty et al.\ \cite{DSB96,SMDB98} proposed the
following BGK-like kinetic model for the \textit{elastic} Enskog
equation:
\beqa
J_E^{(1)}[f]&\to& -\nu_0\left(f-f_0\right)-\frac{f_0}{nT}\left[\mathbf{V}\nabla:\mathsf{P}_c^{(1)}\right.\nn
&&+\left(\frac{mV^2}{dT}-1\right)\left(\nabla\cdot \mathbf{q}_c^{(1)}+\mathsf{P}_c^{(1)}:\nabla\mathbf{u}\right)\nn
&&\left.-\mathsf{A}^{(1)}:\mathsf{D}(\mathbf{V})
-\mathbf{B}^{(1)}\cdot\mathbf{S}(\mathbf{V})\right],
\label{n6.24}
\eeqa
where
\beq
\mathsf{A}^{(1)}\equiv \frac{1}{2T}\int\dd\mathbf{v}\,\mathsf{D}(\mathbf{V})J_E^{(1)}[f_0],
\label{n6.25}
\eeq
\beq
\mathbf{B}^{(1)}\equiv \frac{2m}{(d+2)T^2}\int\dd\mathbf{v}\,\mathbf{S}(\mathbf{V})J_E^{(1)}[f_0].
\label{n6.26}
\eeq
In Eqs.\ (\ref{n6.24})--(\ref{n6.26}), $\mathsf{D}(\mathbf{V})$ and
$\mathbf{S}(\mathbf{V})$ are given by Eqs.\ (\ref{n3.2}) and
(\ref{n3.3}), respectively. Now, making use of (\ref{n6.20}), the
kinetic model (\ref{n6.24}) is extended to inelastic collisions as
\beqa
J_E^{(\alpha)}[f]&\to& -\frac{1+\alpha}{2}\nu_0\left(f-f_0\right)+\frac{\zeta_0}{2}\frac{\partial}{\partial\mathbf{v}}\cdot\left(\mathbf{V}f\right)\nn
&&
-\frac{f_0}{nT}\left[\mathbf{V}\nabla:\mathsf{P}_c^{(\alpha)}+\left(\frac{mV^2}{dT}-1\right)\right.\nn
&&\times\left(\nabla\cdot \mathbf{q}_c^{(\alpha)}+\mathsf{P}_c^{(\alpha)}:\nabla\mathbf{u}\right)\nn
&&\left.-\mathsf{A}^{(\alpha)}:\mathsf{D}(\mathbf{V})
-\mathbf{B}^{(\alpha)}\cdot\mathbf{S}(\mathbf{V})\right],
\label{n6.27}
\eeqa
where $\mathsf{A}^{(\alpha)}=\frac{1}{2}(1+\alpha)\mathsf{A}^{(1)}$,
$\mathbf{B}^{(\alpha)}=\frac{1}{2}(1+\alpha)\mathbf{B}^{(1)}$. The
kinetic model (\ref{n6.27}) for the inelastic Enskog equation turns
out to be  a simplified version of the model introduced in Ref.\
\cite{BDS99}.

\section{Inelastic Maxwell models\label{sec5.3}}
\subsection{Transport coefficients}
Now we return to a single dilute gas. In principle, the same
philosophy behind Eq.\ (\ref{b4}) can be applied to inelastic
Maxwell models (IMM). In that case, the collision operator is
\cite{S03,EB02}
\begin{eqnarray}
&&J^{(\alpha)}[f,f]=\frac{(d+2)\nu_0}{2n\Omega_d}\int \dd{\bf v}_{1}\int \dd\widehat{\bm{\sigma}}\, \left[ \alpha ^{-1}f({\bf r},{\bf v}'')\right.
\nonumber\\
&&\left.\times f({\bf r},{\bf v}
_{1}'')-f({\bf r},{\bf v})f({\bf r},{\bf v}_{1})\right] .
\label{4.1}
\end{eqnarray}
This collision operator verifies again Eq.\ (\ref{3.4}) but now the
cooling rate is exactly given by $\zeta=\zeta_0$, Eq.\ (\ref{3.14}).
The transport coefficients associated with the stress tensor and the
heat flux have the structure of Eqs.\ (\ref{c5})--(\ref{c7}), except
that now the collision frequencies are \cite{S03}
\begin{equation}
\nu_\eta^*=\frac{(1+\alpha)(d+1-\alpha)}{2d}
\label{4.2}
\end{equation}
\begin{equation}
\nu_\lambda^*=\frac{1+\alpha}{d}\left[\frac{d-1}{2}+\frac{1}{8}(d+8)(1-\alpha)\right],
\label{4.3}
\end{equation}
and the kurtosis of the homogeneous cooling state is
\begin{equation}
a_2^\hcs=
\frac{6(1-\alpha)^2}{4d-7+3\alpha(2-\alpha)}.
\label{4.4}
\end{equation}

Using the property \cite{G03b}
\beq
\int\dd\mathbf{v}\,\mathbf{v}J^{(\alpha)}[f_\tg,f_\hcs]=-\frac{d+2}{2d}\nu_0\mathbf{j}_\tg
\label{n4.1}
\eeq
it is straightforward to prove that the self-diffusion coefficient is given by Eq.\ (\ref{n3.6}), except that
\beq
\nu_D^*=\frac{d+2}{2d}\frac{1+\alpha}{2}.
\label{n4.2}
\eeq
The Navier--Stokes transport coefficients characterizing an IMM mixture have been recently derived \cite{GA04}.

\subsection{Model of frictional elastic Maxwell particles}
In the case of \textit{frictional} elastic Maxwell models (EMM), one
makes the replacement (\ref{b4}). The corresponding transport
coefficients have the same forms as for EHS, so they are given again
by Eqs.\ (\ref{c13})--(\ref{c15}) and (\ref{n3.11}). In the same
spirit as before, if we want to optimize the agreement between the
EMM and IMM transport coefficients, the three possible choices for
$\beta$ are
\begin{equation}
\beta_\eta=\nu_\eta^*-\zeta_0^*=\frac{(1+\alpha)^2}{4} ,
\label{4.5}
\end{equation}
\begin{equation}
\beta_\lambda=\frac{d}{d-1}\left(\nu_\lambda^*-\frac{3}{2}\zeta_0^*\right)=\frac{(1+\alpha)^2}{4},
\label{4.6}
\end{equation}
\begin{equation}
\beta_D=\frac{2d}{d+2}\nu_D^*=\frac{1+\alpha}{2}.
\label{n4.3}
\end{equation}
Interestingly, the shear viscosity and the thermal conductivity
routes give consistently  the same expression for $\beta$, this
expression being the square of the self-diffusion value. Since
$a_2^\hcs$, which vanishes for EMM, does not appear either in the
shear viscosity  or in the self-diffusion coefficient of IMM, the
EMM model reproduces exactly those coefficients if
$\beta=\beta_\eta=(1+\alpha)^2/4$ and $\beta=\beta_D=(1+\alpha)/2$,
respectively. However, even if $\beta=\beta_\lambda=(1+\alpha)^2/4$,
the transport coefficients $\lambda$ and $\mu$ for IMM differ from
those for EMM due to the explicitly appearance of $a_2^\hcs$ in the
former case.

\begin{figure}
 \includegraphics[width=.90 \columnwidth]{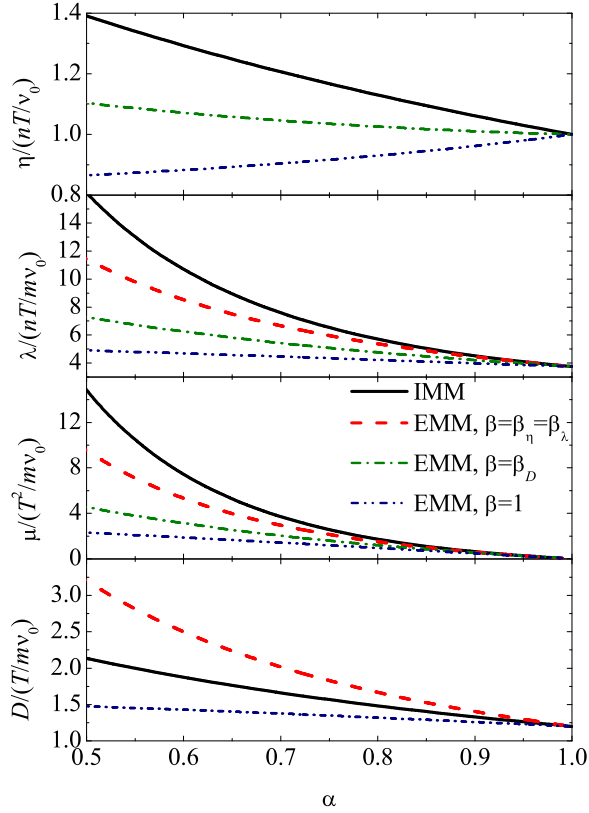}
\caption{(Color online)  Plot of the (reduced) shear viscosity
$\eta/(nT/\nu_0)$, thermal conductivity $\lambda/(nT/m\nu_0)$,
transport coefficient $\mu/(T^2/m\nu_0)$, and self-diffusion
coefficient $D/(T/m\nu_0)$ for three-dimensional  IMM (---) and the
``equivalent'' system of EMM  with $\beta=\beta_\eta=\beta_\lambda$
(-- -- --), $\beta=\beta_D$ (-- $\cdot$ -- $\cdot$ --), and
$\beta=1$ (-- $\cdot\cdot$ -- $\cdot\cdot$ --) as functions of the
coefficient of restitution $\alpha$. Note that in the top panel the
IHS curve and the EHS curve with $\beta=\beta_\eta=\beta_\lambda$
are identical. The same happens in the bottom panel between the IHS
curve and the EHS curve with $\beta=\beta_D$.\label{coeff_IMM}}
 \end{figure}
As Fig.\ \ref{coeff_IMM} shows, the choice
$\beta=\beta_D=(1+\alpha)/2$ describes the behavior of $\eta$,
$\lambda$, and $\mu$ only at a rough qualitative level. The
alternative choice $\beta=\beta_\eta=\beta_\lambda=(1+\alpha)^2/4$
strongly overestimates the diffusion coefficient and gives values
for $\lambda$ and $\mu$ in reasonable agreement with the true IMM
values only for small or moderate inelasticity. Therefore, the
Navier--Stokes transport coefficients of IMM are very poorly
described by the EMM model with a unique expression for $\beta$, in
contrast to what happens in the case
$\text{IHS}\leftrightarrow\text{EHS}$. This a consequence of the
fact that the influence of inelasticity is much stronger in IMM than
in IHS, as reflected on the kurtosis $a_2^\hcs$ and on the transport
coefficients $\eta$, $\lambda$, $\mu$, and $D$ (compare Figs.\
\ref{coeff} and \ref{coeff_IMM}). As a matter of fact, the
coefficients $\lambda$ and $\mu$ diverge at $\alpha=(4-d)/3d$ for
$d=2$ and $d=3$ \cite{S03}, thus indicating the failure of a
hydrodynamic description for IMM with $\alpha\leq(4-d)/3d$.

While, with the exception of the self-diffusion coefficient
\cite{GA04}, the Navier--Stokes transport coefficients of IHS and
IMM differ significantly \cite{S03}, those of frictional EHS and EMM
agree each other [see Eqs.\ (\ref{c13})--(\ref{c15}) and
(\ref{n3.11})], provided the former are evaluated in the first
Sonine approximation and the same value of $\beta$ is used in both
approaches. As a matter of fact, hard spheres and  Maxwell particles
are known to exhibit similar rheological properties in the elastic
case, even in states far from equilibrium \cite{GS03}. This shows
that the EMM system with $\beta=(1+\alpha)/2$ can be usefully
exploited as a model of IHS rather than as a model of IMM, namely
\beq
J_{\text{IHS}}^{(\alpha)}[f,f]\to
\frac{1+\alpha}{2}J_{\text{EMM}}^{(1)}[f,f]+\frac{\zeta_0}{2}\frac{\partial}{\partial\mathbf{v}}\cdot\left(
\mathbf{V}f\right). \eeq Since the collision operator of EMM is
mathematically more tractable than that of EHS, the mapping
$\text{IHS}\to\text{EMM}$ can be considered as intermediate between
the mappings $\text{IHS}\to\text{EHS}$ and
$\text{IHS}\to\text{kinetic models}$.
\section{Velocity moments in the uniform shear flow\label{appA}}
In this Appendix we derive some expressions for the velocity moments
predicted by the BGK and ES kinetic models for uniform shear flow.
Let us focus on the \textit{isotropic} moments
\beq
\langle V^{2N}\rangle=\frac{1}{n}\int \dd\mathbf{V} V^{2N}f(\mathbf{V}).
\label{nA.1}
\eeq
Insertion of the steady-state solution (\ref{6.10}) yields
\begin{eqnarray}
\langle V^{2N}\rangle
&=&
\frac{\beta\nu_0\text{Pr}}{n}\int_0^\infty \dd s\, \exp\left[-\left(\beta\nu_0\text{Pr}+{N}\zeta_0\right)s\right]\nn
&&\times\int \dd\mathbf{V}  f_R \left(\mathbf{V}\right)
|\mathsf{A}(s)\cdot\mathbf{V}|^{2N},
\label{6.11}
\end{eqnarray}
where $A_{ij}(s)=\delta_{ij}-as\delta_{ix}\delta_{jy}$.
To carry out the velocity integral, we consider the diagonal representation of the matrix $\mathsf{R}^{-1}$, namely
\begin{equation}
\mathsf{S}=\frac{mn}{2}\mathsf{U}\cdot\mathsf{R}^{-1}\cdot\mathsf{U}^{-1},
\label{6.12}
\end{equation}
where $\mathsf{U}$ is a unitary matrix whose expression will be omitted here.
The eigenvalues are
\begin{equation}
S_{1,2}=\frac{m}{2T}\frac{2\text{Pr}(\beta+\zeta_0^*)}{2(\text{Pr}\beta+\zeta_0^*)
\pm(1-\text{Pr})\left[\sqrt{d\zeta_0^*(2\beta+d\zeta_0^*)}\mp d\zeta_0\right]},
\label{6.21}
\end{equation}
\begin{equation}
S_3=\frac{m}{2T}\frac{\text{Pr}(\beta+\zeta_0^*)}{\text{Pr}\beta+\zeta_0^*},
\label{6.22}
\end{equation}
where $S_3$ is $(d-2)$-fold degenerate.

Using (\ref{6.12}), the velocity integral in Eq.\ (\ref{6.11}) becomes
\begin{eqnarray}
&&\int \dd\mathbf{V}  f_R \left(\mathbf{V}\right)|\mathsf{A}(s)\cdot\mathbf{V}|^{2N}
=
n\pi^{-d/2}\left(\det \mathsf{S}\right)^{1/2}\nn
&&\times\int \dd\mathbf{W}  \e^{-\mathsf{S}:\mathbf{W}\mathbf{W}}|\mathsf{A}(s)\cdot\mathsf{U}^{-1}\cdot\mathbf{W}|^{2N},
\label{6.13}
\end{eqnarray}
where we have made the change of variables
\begin{equation}
\mathbf{V}\to\mathbf{W}=\mathsf{U}\cdot\mathbf{V}.
\label{6.14}
\end{equation}
It is convenient to decompose the vector $\mathbf{W}$ as
\begin{equation}
\mathbf{W}=W_x\widehat{\mathbf{x}}+W_y\widehat{\mathbf{y}}+\mathbf{W}_\perp,
\label{6.15}
\end{equation}
so that
\begin{eqnarray}
|\mathsf{A}(s)\cdot\mathsf{U}^{-1}\cdot\mathbf{W}|^{2}&=&W_\perp^2+B_{xx}(s)W_x^2+2B_{xy}(s) W_x W_y\nn
&&+B_{yy}(s)W_y^2,
\label{6.16}
\end{eqnarray}
where
\begin{equation}
B_{xx}(s)=1+\frac{1}{2}a^2 s^2 - as\frac{\sqrt{2\beta}+ as\sqrt{d\zeta_0^*}/2}{\sqrt{2\beta+d{\zeta_0^*}}},
\label{6.17x}
\end{equation}
\begin{equation}
B_{yy}(s)=1+\frac{1}{2}a^2 s^2 + as\frac{\sqrt{2\beta}- as\sqrt{d\zeta_0^*}/2}{\sqrt{2\beta+d{\zeta_0^*}}},
\label{6.17y}
\end{equation}
\begin{equation}
B_{xy}(s)=as\frac{as\sqrt{\beta/2}+\sqrt{d\zeta_0^*}}{\sqrt{2\beta+d\zeta_0^*}}.
\label{6.19}
\end{equation}
Therefore,
\begin{widetext}
\begin{eqnarray}
\int \dd\mathbf{V}  f_R \left(\mathbf{V}\right)|\mathsf{A}(s)\cdot\mathbf{V}|^{2N}
&=&
n\pi^{-d/2}\left(\det \mathsf{S}\right)^{1/2}\Omega_{d-2}
\int_{-\infty}^\infty \dd W_x   \e^{-S_1 W_x^2}\int_{-\infty}^\infty \dd W_y   \e^{-S_2 W_y^2}
\int_{0}^\infty \dd W_\perp W_\perp^{d-3}   \e^{-S_3 W_\perp^2}\nn
&&\times\left[W_\perp^2+B_{xx}(s)W_x^2+2B_{xy}(s) W_x W_y+B_{yy}(s)W_y^2\right]^N.
\label{6.20}
\end{eqnarray}
For any given value of $N$, the integrals over $W_x$, $W_y$, and
$W_\perp$ in Eq.\ (\ref{6.20}), as well as the integral over $s$ in
Eq.\ (\ref{6.11}), can be evaluated analytically. In particular,
\beqa
\langle V^4\rangle&=&\frac{dT^2}{{m^2}\Pra\beta {{(\beta
+{\zeta_0^*})}^2}{{(\Pra\beta +2{\zeta_0^*})}^5}}
\left\{(2+d){{\Pra}^6}{{\beta }^8}+2{{\Pra}^4}\left[1+5(2+d)
\Pra+(3+d){{\Pra}^2}\right]{{\beta }^7}{\zeta_0^*}\right.\nn &&+
{{\Pra}^2}\left[6d+4(2+d)\Pra+6(13+7d){{\Pra}^2}+20
(3+d){{\Pra}^3}+3d{{\Pra}^4}\right]{{\beta }^6}{{{\zeta_0^*}}^2}\nn
&&+
2\Pra\left[3d+4(1+4d)\Pra+2(34+25d)\Pra^2+40(3+d)\Pra^3+15d\Pra^4\right]
{{\beta }^5}{{{\zeta_0^*}}^3}\nn &&+
2\left[3d+13d\Pra+5(8+15d)\Pra^2+80(3+d)\Pra^3+60d\Pra^4\right]{{\beta
}^4}{{{\zeta_0^*}}^4}\nn &&+
2\left[14d-(16-41d)\Pra+2(122+39d)\Pra^2+120d\Pra^3\right]{{\beta
}^3}{{{\zeta_0^*}}^5}\nn &&\left.+
2\left[-16+28d+(112+15d)\Pra+120d\Pra^2\right] {{\beta
}^2}{{{\zeta_0^*}}^6}+ 4\left[8+d(5+24\Pra)\right]\beta
{{{\zeta_0^*}}^7}+18d{{{\zeta_0^*}}^8}\right\},
\label{nA.2}
\eeqa
where use has been made of Eq.\ (\ref{6.7bis}).

Non-isotropic velocity moments can be obtained in a similar way,
although they are more complicated than the isotropic ones. In the
case of the BGK model ($\Pra=1$), one has
\beq
\langle
V_x^{\ell}V_y^{2k-\ell}V_\perp^{2k'}\rangle=\frac{\beta\nu_0}{n}
\int\dd
s\,\exp\left[-\left(\beta\nu_0+(k+k')\zeta_0\right)s\right]\int\dd\mathbf{V}\left(V_x-asV_y\right)^\ell
V_y^{2k-\ell}V_\perp^{2k'}f_0(\mathbf{V}).
\label{nA.3}
\eeq
After expanding $\left(V_x-asV_y\right)^\ell$ and carrying out the integrations over $\mathbf{V}$ and $s$, one finally gets
\begin{eqnarray}
\langle V_x^{\ell}V_y^{2k-\ell}V_\perp^{2k'}\rangle&=&(-1)^\ell \left(\frac{2T}{m}\right)^{k+k'}\frac{\Omega_{d-2}}{2\pi^{d/2}}
\Gamma\left(\frac{d}{2}+k'-1\right)
\beta
\sum_{q=0}^{[\ell/2]}\frac{\ell!}{(2q)!}\Gamma\left(q+\frac{1}{2}\right)
\Gamma\left(k-q+\frac{1}{2}\right)\nn
&&\times
\left[\frac{d\zeta_0^*(\beta+\zeta_0^*)^2}{2\beta}\right]^{\ell/2-q}\left[\beta+(k+k')\zeta_0^*\right]^{-(\ell-2q+1)}.
\label{6.25}
\end{eqnarray}

\end{widetext}

\end{document}